\documentclass[aps,twocolumn,prd,superscriptaddress,nofootinbib,floats,floatfix]{revtex4-1}

\usepackage[T1]{fontenc}
\usepackage[utf8]{inputenc}
\usepackage{graphicx}
\usepackage{xspace}
\usepackage{amsmath}
\usepackage{amsfonts}
\usepackage{amssymb}
\usepackage{latexsym}
\usepackage{microtype}
\usepackage{enumerate}
\usepackage{tabularx}
\usepackage{lmodern}
\usepackage{verbatim}
\usepackage[dvipsnames,x11names,svgnames,rgb,table]{xcolor}
\usepackage[colorlinks]{hyperref}
\usepackage{orcidlink}
\usepackage{physics}

\DeclareMathAlphabet{\mathpzc}{OT1}{pzc}{m}{it}

\definecolor{amethyst}{HTML}{a45ee5}
\definecolor{austriawien}{HTML}{441678}
\definecolor{dodgerblue}{HTML}{1E90FF}
\definecolor{newterms}{HTML}{1E90FF}

\AtBeginDocument{
  \hypersetup{
    citecolor=dodgerblue,
    linkcolor=dodgerblue,   
    urlcolor=dodgerblue,
    filecolor=dodgerblue
    }
}

\newcommand{\calG}{\mathcal{G}}

\newcommand{\bham}{\affiliation{School of Physics and Astronomy, University of Birmingham, Edgbaston, Birmingham, B15 2TT, United Kingdom}}

\newcommand{\igwa}{\affiliation{Institute for Gravitational Wave Astronomy, University of Birmingham, Edgbaston, Birmingham, B15 2TT, United Kingdom}}

\begin{document}

\title{High-Post-Newtonian-Order Dynamics Induced by Tail-of-Tail Interactions: The Non-Geodesic Terms}

\author{Geraint Pratten \orcidlink{0000-0003-4984-0775}}
\email{g.pratten@bham.ac.uk}
\bham 
\igwa

\hypersetup{pdfauthor={G.~Pratten}}
\date{\today}

\begin{abstract}
We compute the tail-of-tail contribution to the conservative dynamics of eccentric, non-spinning compact binaries to relative 1PN order and to $\mathcal O(e_t^{12})$.  
Using the $1$PN quasi-Keplerian dynamics in harmonic coordinates, we derive the Delaunay-averaged Hamiltonian at $5.5$PN and $6.5$PN order and match it to the effective-one-body description, allowing us to determine the corresponding contributions to the non-geodesic EOB $Q$ potential through the $\mathcal{O}(p_r^{12})$, including the dependence on the symmetric mass ratio up to $\mathcal{O}(\nu^2)$.  
The terms linear in the mass ratio reproduce the available first-order self-force results, while the quadratic terms provide qualitatively new eccentric second-order self-force predictions arising from the tail-of-tail terms. 
We independently rederive the averaged Hamiltonian using a Fourier--Bessel decomposition of the hereditary interaction.  
Applying the first law at fixed orbital frequencies, we recover known contributions to the first-order self-force redshift through $\mathcal O(e^{12})$ and obtain the complete tail-of-tail contribution to the second-order inverse redshift at $5.5$PN and $6.5$PN through $\mathcal O(e^{10})$.
\end{abstract}

\maketitle

\section{Introduction}
\label{sec:intro}
In recent years, significant progress on the general-relativistic two-body problem has been driven by the synergy between a diverse set of complementary approaches: post-Newtonian (PN) theory~\cite{Blanchet:2013haa}, post-Minkowskian (PM) calculations and scattering amplitudes~\cite{Damour:2016gwp,Bjerrum-Bohr:2018xdl,Cheung:2018wkq,Kosower:2018adc,Bern:2019crd,Bern:2019nnu}, effective field theory (EFT)~\cite{Goldberger:2004jt,Goldberger:2009qd,Foffa:2013qca,Porto:2016pyg,Foffa:2021pkg}, gravitational self-force (GSF)~\cite{Barack:2018yvs}, and numerical relativity~\cite{Palenzuela:2020tga}. 
Frameworks such as the effective-one-body (EOB) formalism~\cite{Buonanno:1998gg,Buonanno:2000ef} and the Tutti Frutti approach pioneered in~\cite{Bini:2019nra,Bini:2020hmy,Bini:2020nsb,Bini:2025vuk} allow us to combine information from the individual approaches into a complete description of the binary dynamics. 
The EOB formalism plays a particularly central role, as its resummed potentials provide a common framework into which PN, PM, and GSF results can be transcribed. 

These theoretical programs are increasingly driven by observational
demands.
The LIGO--Virgo--KAGRA detectors~\cite{LIGOScientific:2014pky,VIRGO:2014yos,Aso:2013eba} have established compact-binary coalescences as routine observations~\cite{LIGOScientific:2018mvr,LIGOScientific:2020ibl,KAGRA:2021vkt,LIGOScientific:2025slb,LIGOScientific:2026wfs}, and next-generation facilities, such as LISA~\cite{LISA:2024hlh}, the Einstein Telescope~\cite{ET:2025xjr}, and Cosmic Explorer~\cite{Evans:2021gyd}, will deliver signal-to-noise ratios that place stringent accuracy requirements on waveform models~\cite{Purrer:2019jcp,Hu:2022rjq,Kapil:2024zdn}. 
Orbital eccentricity poses a particular challenge in this context, while also serving as one of the most informative tracers of astrophysical binary formation channels~\cite{Zevin:2021rtf,Fumagalli:2024gko}.
Several formation pathways, including dynamical
interactions in dense star clusters and galactic nuclei, secular
evolution in hierarchical multiples, and interactions in AGN
disks, can produce binaries that retain measurable eccentricity in
the sensitive band of ground-based detectors
~\cite{Samsing:2017xmd,Rodriguez:2018pss,Zevin:2018kzq,
Tagawa:2020jnc,Gondan:2020svr,Samsing:2020tda,Trani:2021tan,
DallAmico:2023neb,Stegmann:2025clo}.
There is already mounting evidence for orbital eccentricity in population-level studies~\cite{Gupte:2024jfe,Morras:2026mrv,Pompili:2026yxq} as well as in individual events, such as the neutron star-black hole binary GW200105~\cite{Morras:2025xfu,Planas:2025plq,Kacanja:2025kpr,Jan:2025fps,Phukon:2025cky}.

These developments have spurred rapid progress in waveform models for
eccentric binaries across all major frameworks.
Within the PN framework, ready-to-use inspiral models for precessing-eccentric binaries are now available~\cite{Klein:2021jtd,Arredondo:2024nsl,Morras:2025nlp,Morras:2026fho}.
In the EOB approach, eccentric extensions of both the \texttt{TEOBResumS} and \texttt{SEOBNR} families have reached maturity, with and without spin precession~\cite{Chiaramello:2020ehz,Khalil:2021txt,Ramos-Buades:2021adz,Liu:2023dgl,Gamboa:2024hli,Nagar:2024dzj,Gamba:2024cvy,Nagar:2024oyk,Albanesi:2025txj,Gamboa:2026jht,Lynch:2026ibo}.
Eccentricity has also been recently incorporated into the
phenomenological frameworks~\cite{Planas:2025feq,Ramos-Buades:2026kbq}, building on the quasi-circular foundations introduced in~\cite{Pratten:2020fqn,Garcia-Quiros:2020qpx,Pratten:2020ceb,Estelles:2020osj}. 
All of these models ultimately rest on analytic solutions to the two-body dynamics, with their accuracy determined by the available PN, mass-ratio, and eccentricity information.
Extending the conservative dynamics of eccentric binaries to higher PN orders, including the non-local hereditary sectors, and obtaining self-force information through second order in the mass ratio are  important stepping stones towards improving these models.

Hereditary contributions are of particular interest as they differ qualitatively from instantaneous terms~\cite{Blanchet:2013haa}.
Beginning at 4PN order, the conservative dynamics acquires nonlocal-in-time contributions from gravitational-wave tails, generated by back-scattering off the curvature sourced by the total mass~\cite{ Blanchet:1985sp,Blanchet:1986dk,Blanchet:1987wq,Blanchet:1993ec,Blanchet:1997jj,Foffa:2011np,Damour:2014jta,Galley:2015kus}.
Starting at $5.5$PN order, non-local-in-time interactions mediated by tail-of-tail effects also contribute to the conservative dynamics~\cite{Shah:2013uya,Bini:2013rfa,Blanchet:2013txa,Bini:2020wpo,Bini:2025zvk}.
Recent work extended the tail-of-tail analysis to eccentric motion by evaluating the time-split action through relative 1PN order and $\mathcal O(e_t^2)$~\cite{Bini:2025zvk}.
The resulting averaged dynamics determines the corresponding $5.5$PN and $6.5$PN corrections to the EOB potentials $A$ and $\bar D$, including terms at $\mathcal O(\nu^2)$~\cite{Bini:2016qtx,Bini:2020wpo,Bini:2025zvk}. 
This truncation, nevertheless, leaves the non-geodesic contributions to the EOB Hamiltonian undetermined, as the $Q(u,p_r;\nu)$ potential in the Damour--Jaranowski--Sch\"afer (DJS) gauge~\cite{Damour:2008qf} only contributes at quartic order in the radial momentum and therefore first contributes at quartic order in eccentricity.

The aim of the present work is to extend the calculations presented in~\cite{Bini:2020wpo,Bini:2025zvk} to higher orders in eccentricities, allowing us to determine the corrections to the non-geodesic EOB $Q$ potential. 
Working in harmonic coordinates with 1PN-accurate quasi-Keplerian (QK) orbital dynamics, we extend the Delaunay-averaged tail-of-tail Hamiltonian of Ref.~\cite{Bini:2025zvk} through $\mathcal O(e_t^{12})$ at relative 1PN order. 
Matching to the Delaunay averaged EOB Hamiltonian order-by-order in eccentricity allows us to determine the tail-of-tail contributions to the $Q$ potential through the $\mathcal{O}(p_r^{12})$, at $5.5$PN and $6.5$PN, including the complete $O(\nu^2)$ dependence. 
The $O(\nu)$ terms reproduce the known GSF-derived coefficients through $p_r^8$~\cite{Bini:2016qtx,Bini:2020wpo}, while the higher-order $p_r$ terms and the $O(\nu^2)$ parts constitute new predictions.

We validate our results through three complementary checks.
First, we rederive the averaged Hamiltonian by decomposing the source multipoles into discrete Fourier modes, and then evaluate the hereditary kernels mode-by-mode following the spectral methods of Refs.~\cite{Arun:2007rg,Arun:2007sg,Damour:2015isa,
Loutrel:2016cdw,Trestini:2025yyc}.
Second, we use the first law of binary black hole mechanics~\cite{LeTiec:2011ab,LeTiec:2015kgg}, generalised to hereditary interactions in Ref.~\cite{Blanchet:2017rcn}, to compute the tail-of-tail contribution to the averaged redshift at fixed orbital frequencies.
At first order in the mass ratio, the resulting expression agrees through $\mathcal O(e^{12})$ with the independent GSF calculation of Ref.~\cite{Munna:2022gio}.
Third, a complementary EOB calculation provides a self-consistency check of the second-order self-force ($2$SF) terms through $\mathcal O(e^2)$.
As a further result, we obtain the complete half-integer-PN
contributions to the $2$SF inverse redshift at $5.5$PN and $6.5$PN through $\mathcal O(e^{10})$.
These results provide explicit weak-field benchmarks for ongoing extensions of second-order self-force calculations~\cite{Gralla:2012db,Pound:2014koa,Pound:2017psq,Warburton:2021kwk,Wardell:2021fyy,Mathews:2021rod,Miller:2023ers,Upton:2025bja}.

The paper is organised as follows. Section~\ref{sec:qk} summarises the
quasi-Keplerian parameterisation and our conventions.
Section~\ref{sec:tailoftails} introduces the time-split tail-of-tail
action and derives the Delaunay-averaged Hamiltonian through
$\mathcal O(e_t^{12})$. 
Section~\ref{sec:delEOB} performs the EOB matching and presents the $Q$-potential coefficients.
Section~\ref{sec:spectral} develops the spectral cross-validation, and Sec.~\ref{sec:redshift} derives the first- and second-order self-force redshift contributions.
Throughout, we use units $G=c=1$, except where $\eta=1/c$ is retained as a PN bookkeeping parameter. 
The total mass is denoted $M = m_1 + m_2$, with $\mu = m_1 m_2 / M$ the reduced mass, and $\nu=\mu/M$ the symmetric mass ratio.

\section{Quasi-Keplerian Orbits}
\label{sec:qk}
We will work with elliptic-like orbits using the QK parameterisation in harmonic coordinates at $1$PN order accuracy~\cite{Damour:1985der}. 
Following ~\cite{Bini:2025zvk}, we adopt mass-rescaled dimensionless orbital variables, such that $r = r^{\rm phys} / (G M)$, $a_r = a_r^{\rm phys} / (G M)$, $n = n^{\rm phys} (G M / c^3)$.
We use $M = m_1 + m_2$, $\mu = m_1 m_2 / M$, and $\nu = \mu / M$ for the total mass, reduced mass, and symmetric mass ratio respectively.
At this order, the QK parameterisation reduces to~\cite{Damour:1985der,Damour:1986der2}
\begin{align}
\label{eq:Kepler}
n t &= \ell = u - e_t \sin u, \\
r &= a_r \left( 1 - e_r \cos u \right), \\
\phi &= 2 K \arctan \left[ \left( \frac{1 + e_{\phi}}{1 - e_{\phi}} \right)^{1/2} \tan \frac{u}{2} \right],
\end{align}
where $n = 2\pi/P$ with $P$ the period of the radial motion, $\ell$ is the mean anomaly, $u$ is the eccentric anomaly, $K = 1+k$ encodes the periastron advance with $k$ the fractional periastron advance per radial period, and $a_r$, $e_t$, $e_r$, and $e_{\phi}$ are the semi-major axis and the time, radial, and angular eccentricities, respectively. 
Unless otherwise stated, all QK elements in this section are defined in harmonic coordinates. 
The true anomaly $v(u)$ can be related to the eccentric anomaly $u$ through~\cite{Damour:1985der}
\begin{equation}
\label{eq:vtanom}
v = 2 \arctan \!\left[
\left( \frac{1 + e_{\phi}}{1 - e_{\phi}} \right)^{1/2}
\tan \frac{u}{2}
\right],
\end{equation}
such that
\begin{align}
\cos v &= \frac{\cos u - e_{\phi}}{1 - e_{\phi}\cos u}, \\
\sin v &= \frac{\sqrt{1-e_{\phi}^2}\,\sin u}{1 - e_{\phi}\cos u}.
\end{align}
At $1$PN, the relation between the mean motion and the radial semi-major axis is given by
\begin{align}
n = a_r^{-3/2} \left[ 1 + \frac{1}{2}\frac{\eta^2}{a_r}(-9 + \nu) \right],
\end{align}
where we have introduced $\eta = 1/c$ to track the PN order~\cite{Bini:2025zvk}, with $\eta^2$ denoting terms at $1$PN order. 
It will also be useful to describe the orbit in terms of the reduced binding energy and angular momentum~\cite{Damour:1988mr,Arun:2007sg,LeTiec:2015kgg,Bini:2020wpo},
\begin{equation}
\varepsilon = -\frac{2E_{\rm b}}{\mu c^2},
\qquad
j = \varepsilon \left(\frac{c L}{G M \mu}\right)^2 .
\label{eq:x-j-definitions}
\end{equation}
Here, $E_b$ denotes the conservative binding energy and $L$ the conservative orbital angular momentum. 
The QK orbital elements can be expressed in terms of $\varepsilon$ and $j$ as~\cite{Damour:1985der,Damour:1988mr,Bini:2020wpo}
\begin{align}
    a_r &= \frac{1}{\varepsilon} \left[ 1 - \frac{7-\nu}{4} \eta^2 \varepsilon \right], \label{eq:QK_ar}
    \\
    n &= \varepsilon^{3/2} \left[ 1 - \frac{15-\nu}{8} \eta^2 \varepsilon \right], \label{eq:QK_n}
    \\
    e_t^2 &= (1-j)+\eta^2 \varepsilon \left[ \frac{17-7\nu}{4}j -2 + 2\nu \right], \label{eq:QK_et}
    \\ 
    e_r^2 &= (1-j) + \eta^2 \varepsilon \left[ 6 - \nu - \frac{15-5\nu}{4} j \right], \\
    e^2_{\phi} &= (1-j) + \eta^2 \varepsilon \left[ 6 - \frac{15-\nu}{4} j \right], \\
    K &= 1 + k,
\end{align}
where $k = K - 1 = \Omega_{\phi} / \Omega_r - 1$ is the fractional periastron advance per radial period~\cite{Damour:1988mr,Memmesheimer:2004cv}, and up to $2$PN is given by~\cite{Damour:1988mr}
\begin{align}
k &=\frac{3 \eta^2 \varepsilon}{j} \left\lbrace 1 + \eta^2 \varepsilon \left[-\frac{5 - 2 \nu}{4} + \frac{35 - 10 \nu}{4 j} \right] \right\rbrace.
\end{align}
Although the orbital dynamics used here are only required through $1$PN, we retain the 2PN correction to $k$ as this will be necessary in Sec.~\ref{sec:2SF-eccentric}, when calculating the second-order self-force correction to the redshift invariants. 
The energy and angular momentum can be equivalently written in terms of the QK orbital elements,
\begin{align}
    \varepsilon &= \frac{1}{a_r} - \frac{7 - \nu}{4 a_r^2} \eta^2, \\
    j &= (1 - e_t^2) + \frac{\eta^2}{a_r} \left[ 4 + 2(\nu - 3) e^2_t - \frac{7 - \nu}{4} (1 - e^2_t) \right] .
\end{align}
We will use these relations repeatedly when converting between harmonic QK variables, EOB variables, and fixed-frequency orbital parameters.

\section{The Tail-of-Tail Action}
\label{sec:tailoftails}
We now collect the ingredients required to construct the conservative tail-of-tail Hamiltonian through relative 1PN order. 
Where possible, our notation follows Ref.~\cite{Bini:2025zvk}, and we explicitly state the sign and principal-value conventions adopted in this work.
We work with the time-split tail-of-tail action~\cite{Bini:2020hmy,Bini:2020wpo,Bini:2025zvk}
\begin{align}
S^{{\rm tt}}_{\rm time\text{-}split} &= -\frac14  \left(\frac{G{\mathcal M}}{c^3}\right)^2G \int dt \int _{-\infty}^\infty \frac{d \tau}{\tau} {\mathcal G}^{\rm split}(t,\tau)\,,\qquad
\end{align}
where
\begin{align}
{\calG}^{\rm split}(t,\tau) &= \sum_{l\ge 2}\frac{1}{c^{2l+1}}\left( a_l \beta_l^{\rm (even)}{\mathcal G}_I^l(t,\tau)\right.\nonumber\\
&\qquad + \left. \frac{b_l}{c^2} \beta_l^{\rm (odd)}{\mathcal G}_J^l(t,\tau)\right).
\end{align}
Here, we have introduced the $1$PN accurate ADM mass,
\begin{align}
\mathcal{M} &= M \left( 1 - \eta^2 \frac{\nu}{2 a_r} \right),
\end{align}
which will contribute to the $\nu$-dependent part of the relative-$1$PN tail-of-tail Hamiltonian.  

The kernels $\calG_{X}^{l}$ are defined by
\begin{align}
\calG_I^{\,l}(t,\tau)
&\equiv
I_L^{(l+1)}(t)
\Bigl[
    I_L^{(l+2)}(t+\tau)
    -
    I_L^{(l+2)}(t-\tau)
\Bigr],
\\
\calG_J^{\,l}(t,\tau)
&\equiv
J_L^{(l+1)}(t)
\Bigl[
    J_L^{(l+2)}(t+\tau)
    -
    J_L^{(l+2)}(t-\tau)
\Bigr],
\end{align}
where $I_L$ and $J_L$ are the MPM Blanchet-Damour-Iyer multipole moments~\cite{Blanchet:1985sp,Blanchet:1987wq,Blanchet:1992br,Damour:1990gj}, and the superscripts denote the number of time derivatives applied to the multipole moment. 
This can be expressed using the short-hand notation introduced in~\cite{Bini:2025zvk},
\begin{equation}
\label{eq:Stot}
\mathsf{G}_X^l = \int dt \int_{-\infty}^{\infty} \frac{d\tau}{\tau}\,\mathcal{G}_X^l(t,\tau),
\end{equation}
which results in the following time-split action
\begin{align}
S_{\rm time\text{-}split}^{\rm tt}
&=
-\frac{G}{4}
\left(\frac{G\mathcal{M}}{c^3}\right)^2
\sum_{l\ge2}
\frac{1}{c^{2l+1}}
\Biggl[
    a_l \beta_l^{\rm (even)} \mathsf{G}_I^l
\nonumber\\
&\qquad\qquad
    +
    \frac{b_l}{c^2}
    \beta_l^{\rm (odd)}
    \mathsf{G}_J^l
\Biggr].
\label{eqtailoftail2}
\end{align}
Through relative $1$PN order, only the $\mathsf{G}^2_I$, $\mathsf{G}^3_I$, and $\mathsf{G}^2_J$ sectors contribute. 
The mass quadrupole is required through $1$PN, whereas the mass octupole and current quadrupole are only required at Newtonian order, as their leading order contributions are suppressed by an additional PN order.  
The multipole and beta coefficients are given by~\cite{Blanchet:1987wq,Blanchet:1997jj,Goldberger:2009qd,Almeida:2021jyt,Fucito:2024wlg,Ivanov:2025ozg}
\begin{align}
a_2 = \frac{1}{5}, \; a_3 = \frac{1}{189}, \; b_2 = \frac{16}{45}, \;
\beta_2 = - \frac{214}{105}, \; \beta_3 = - \frac{26}{21},
\end{align}
where $\beta_{l}^{\rm even} = \beta_{l}^{\rm odd}$~\cite{Fucito:2024wlg,Ivanov:2025ozg,Bini:2025vuk}.
Evaluating these contributions requires only the source mass quadrupole $I_{ab}$, mass octupole $I_{abc}$, and current quadrupole $J_{ab}$. 
In harmonic coordinates at $1$PN order, these moments are given by~\cite{Blanchet:1989cu,Blanchet:2013haa}
\begin{align}
\label{eq:1PN_multipoles}
I_{ij}
&= \mu r_{\langle ij \rangle}
   \left[
      1 + \frac{29}{42} \eta^2(1-3\nu) v^2 
      -\frac{(5-8\nu)}{7} \eta^2\frac{GM}{r}
   \right]
   \nonumber\\
&\quad
   +\mu\eta^2 \frac{1-3\nu}{21}
   \left[
      -12(\mathbf v\cdot\mathbf r)\,
      r_{\langle i}v_{j\rangle}
      +11r^2v_{\langle ij\rangle}
   \right], \nonumber\\
I_{ijk}
&= \mu\sqrt{1-4\nu}\,r_{\langle ijk\rangle},
\nonumber\\
J_{ij}
&= \mu\sqrt{1-4\nu}\,
   \eta_{kl\langle i}r_{j\rangle k}v_l,
\end{align}
where
\begin{align}
r_{\langle ijk\rangle} &= r_{ijk} -\frac{3}{5}r^2 \delta_{(ij} r_{k)} ,
\nonumber\\
\eta_{kl\langle i}r_{j\rangle k}v_l &= (\mathbf r\times\mathbf v)_{\langle i}r_{j \rangle}.
\end{align} 
Here, we use standard notation, notably $T^{\langle ij \rangle}$ for the symmetric trace-free part of a tensor, $T^{(ij)} = (T^{ij} + T^{ji})/2$ for the symmetric part, $T_{ij\cdots} = T_i T_j \cdots$ for tensorial products, and $\eta_{ijk}$ is the Levi-Civita symbol.

From the action,
\begin{align}
S^{{\rm tt}}_{\rm time\text{-}split} &= - \int dt \, H^{{\rm tt}} (t),
\end{align}
the time-split Hamiltonian can be expressed as
\begin{align}
\label{eq:Htot}
H^{\rm tt}(t) &= H^{\rm tt}_{I_2} (t) \nonumber \\
&\quad + \eta^2 \left[ H^{\rm tt}_{I_3} (t) + H^{\rm tt}_{J_2} (t) \right],
\end{align}
where the relative-1PN correction to $I_3$ and $J_2$ sectors has been explicitly factored out. 
The contributions to the tail-of-tail Hamiltonian can be written in a compact form in terms of principal-value (PV) integrals by introducing the following short-hand notation~\cite{Bini:2025zvk}
\begin{equation}
\label{eq:HilbertPV}
H\!\left[X_L^{(n,m)}(t)\right]
\equiv
X_L^{(n)}(t)
\operatorname{PV}\!\int_{-\infty}^{\infty}
\frac{dt'}{t'-t}\,
X_L^{(m)}(t').
\end{equation}
With the convention adopted here, the orbit-averaged bilinear is the negative of a positive-definite spectral norm, as discussed in Sec.~\ref{sec:spectral}.
The contributions to the tail-of-tail Hamiltonian can now be compactly expressed as~\cite{Bini:2025zvk}
\begin{align}
H_{I_2}^{\rm tt}(t)
&=
-\frac{G}{4}
\left( \frac{G\mathcal{M}}{c^3} \right)^2
\frac{1}{c^5}
\frac{214}{525}
\, 2 H\!\left[I_2^{(3,4)}(t)\right],
\\
H_{J_2}^{\rm tt}(t)
&=
-\frac{G}{4}
\left( \frac{G\mathcal{M}}{c^3} \right)^2
\frac{1}{c^5}
\frac{3424}{4725}
\,2 H\!\left[J_2^{(3,4)}(t)\right],
\\
H_{I_3}^{\rm tt}(t)
&=
-\frac{G}{4}
\left( \frac{G\mathcal{M}}{c^3} \right)^2
\frac{1}{c^5}
\frac{26}{3969}
\, 2 H\!\left[I_3^{(4,5)}(t)\right].
\end{align}
Each contribution is constructed from bilinears in the source multipole moments, which we jointly PN-expand and eccentricity-expand. 
We retain the mass-quadrupole through relative $1$PN order, and all bilinears through $\mathcal{O}(e_t^{12})$. 
This will be sufficient to determine the $\mathcal{O}(p^{12}_r)$ contributions to the EOB $Q$ potential, as shown in Sec.~\ref{sec:delEOB}. 

Schematically, the bilinears of interest have the form
\begin{align}
\label{eq:I2bilinear}
I_2^{(3,4)}(u,u')
&=
I_{ij}^{(3)}(t)\,
I_{ij}^{(4)}(t')
\Big|_{t=t(u),\,t'=t(u')},
\\
\label{eq:J2bilinear}
J_2^{(3,4)}(u,u')
&=
J_{ij}^{(3)}(t)\,
J_{ij}^{(4)}(t')
\Big|_{t=t(u),\,t'=t(u')},
\\
\label{eq:I3bilinear}
I_3^{(4,5)}(u,u')
&=
I_{ijk}^{(4)}(t)\,
I_{ijk}^{(5)}(t')
\Big|_{t=t(u),\,t'=t(u')},
\end{align}
which, when PN and eccentricity-expanded, lead to a common structure of the form
\begin{align}
X(u,u')
&=
\sum_{p=0}^{p_{\rm max}}
e_t^p \left[ X_{(\eta^0,e_t^p)}(u,u')
+ \eta^2 X_{(\eta^2,e_t^p)}(u,u') \right].
\end{align}
The terms up to $\mathcal{O}(e^2_t)$ have already been presented in~\cite{Bini:2025zvk}, and are in full agreement with the results presented here. 
Following~\cite{Bini:2025zvk}, we evaluate Eq.~\eqref{eq:HilbertPV} after changing integration variable from $dt' / (t' - t)$ to $du'$, $u'$, and $u$, leading to a finite set of principal-value integrals. 
Representative bilinears through $\mathcal{O}(e^4_t)$ are shown in Tables~\ref{tab:1}-\ref{tab:3} in App.~\ref{app:bilinears}. 
The complete expressions through $\mathcal{O}(e^{12}_t)$ are given in the ancillary material.

\subsection{Delaunay Averaged Hamiltonian}
We define the Delaunay average over one radial period by
\begin{align}
\langle H_{X_L}^{\rm{tt}} \rangle &\equiv \frac{\displaystyle\oint H_{X_L}^{\rm{tt}}(t)\,dt}{\displaystyle\oint dt}.
\end{align}
The radial period entering the denominator is defined through relative $1$PN order by
\begin{align}
\displaystyle\oint dt &= 2\pi a_r^{3/2} + a_r^{1/2} \pi \eta^2 \left( 9 - \nu \right),
\end{align}
\noindent
which follows directly from $P = 2\pi / n$ and the $1$PN relation in Eq.~\eqref{eq:QK_n}. 

Applying the Delaunay averaging procedure to the multipole moment bilinears allows us to determine all the required contributions to construct the averaged tail-of-tail Hamiltonian.
For the mass quadrupole $I_2$, we have two contributions at fractional $1$PN order
\begin{equation}
\left\langle H\!\left[I_2^{(3,4)}\right]\right\rangle
=
\left\langle H\!\left[I_{2,\eta^0}^{(3,4)}\right]\right\rangle
+\eta^2
\left\langle H\!\left[I_{2,\eta^2}^{(3,4)}\right]\right\rangle,
\end{equation}
where the two terms are given by
\begin{widetext}
\begin{align}
\label{eq:HI2N}
2\left\langle H \left[I_{2,\eta^0}^{(3,4)}(t)\right]\right\rangle
&= \frac{\pi\nu^2}{a_r^{13/2}}\bigg[
  -128 - \frac{4670}{3}e_t^2 - \frac{42955}{6}e_t^4 - \frac{6204647}{288}e_t^6
  - \frac{352891481}{6912}e_t^8 \nonumber\\
&\qquad
  - \frac{286907786543}{2764800}e_t^{10} - \frac{6287456255443}{33177600}e_t^{12}
  \bigg],
\end{align}
\begin{align}
\label{eq:HI21PN}
2\left\langle H\left[I_{2,\eta^2}^{(3,4)}(t)\right]\right\rangle
&= \frac{\pi\nu^2}{a_r^{15/2}} \Bigg[ \bigg(
  \frac{31168}{21} + \frac{256805}{21}e_t^2 + \frac{339379}{12}e_t^4
  - \frac{309479}{1728}e_t^6 \nonumber\\
&\qquad
  - \frac{19714929593}{96768}e_t^8 - \frac{1527310235551}{1843200}e_t^{10}
  - \frac{351712023425809}{154828800}e_t^{12} \bigg) \nonumber\\
&\quad + \bigg(
  -\frac{4288}{7} - \frac{158993}{21}e_t^2 - \frac{146445}{4}e_t^4
  - \frac{67251767}{576}e_t^6 \nonumber\\
&\qquad
  - \frac{28402847971}{96768}e_t^8 - \frac{388346663887}{614400}e_t^{10}
  - \frac{566906059048153}{464486400}e_t^{12} \bigg) \nu \Bigg] .
\end{align}
\end{widetext}
Similarly, for the current quadrupole $J_2$ and mass octupole $I_3$ respectively, we find
\begin{widetext}
\begin{align}
\label{eq:HavJ2}
2\left\langle H\left[J_2^{(3,4)}(t)\right]\right\rangle
={}& -(1-4\nu)\,\frac{\pi\nu^2}{a_r^{15/2}}\bigg[
  1 + 30\,e_t^2 + \frac{7321}{32}e_t^4 + \frac{279485}{288}e_t^6 \nonumber\\
&\quad + \frac{73162261}{24576}e_t^8 + \frac{3426964009}{460800}e_t^{10}
  + \frac{105664935873}{6553600}e_t^{12}\bigg] , \\
2\left\langle H\left[I_3^{(4,5)}(t)\right]\right\rangle
={}& -(1-4\nu)\,\frac{\pi\nu^2}{a_r^{15/2}}\bigg[
  \frac{49209}{5} + \frac{910188}{5}e_t^2 + \frac{367780251}{320}e_t^4
  + \frac{849097483}{192}e_t^6 \nonumber\\
&\quad + \frac{104770539213}{8192}e_t^8 + \frac{47178621858259}{1536000}e_t^{10}
  + \frac{25457292864087389}{393216000}e_t^{12}\bigg] ,
\label{eq:HavI3}
\end{align}
\end{widetext}
Through $\mathcal{O}(e^2_t)$, these expressions agree with the results presented in Ref.~\cite{Bini:2025zvk}. 
Their overall signs are independently fixed by the spectral norm representation that we introduce in Sec.~\ref{sec:spectral}. 
In particular, this identifies an overall sign typo in the $I_3$ contribution quoted in Ref.~\cite{Bini:2025zvk}, noting that this does not affect the final results. 

Combining the three multipolar terms, including the $1$PN correction to $\mathcal{M}^2$, yields the Delaunay-averaged tail-of-tail Hamiltonian in harmonic coordinates
\begin{widetext}
\begin{align}
\label{eq:HNtotharmonic}
\left\langle H^{\rm tt} \right\rangle_{\eta^0}
&=
\frac{\pi \nu^2}{a_r^{13/2}}
\Bigg[
\frac{6848}{525}
+\frac{49969}{315}e_t^2
+\frac{919237}{1260}e_t^4
+\frac{663897229}{302400}e_t^6
\nonumber\\
&\qquad
+\frac{37759388467}{7257600}e_t^8
+\frac{4385590451443}{414720000}e_t^{10}
+\frac{672757819332401}{34836480000}e_t^{12}
\Bigg],
\end{align}
\begin{align}
\label{eq:H1PNtotharmonic}
\left\langle H^{\rm tt} \right\rangle_{\eta^2}
&=
-\frac{\pi \nu^2}{a_r^{15/2}}
\Bigg[
\Bigg(
\frac{991861}{7350}
+\frac{593849}{630}e_t^2
+\frac{54097493}{56448}e_t^4
-\frac{3935427469}{529200}e_t^6
\nonumber\\
&\qquad
-\frac{4578610749551}{108380160}e_t^8
-\frac{368735424136853}{2709504000}e_t^{10}
-\frac{1062619266017932697}{3121348608000}e_t^{12}
\Bigg)
\nonumber\\
&\qquad
+\Bigg(
\frac{4982}{315}
+\frac{1473251}{2450}e_t^2
+\frac{1655819311}{352800}e_t^4
+\frac{253642801109}{12700800}e_t^6
\nonumber\\
&\qquad\qquad
+\frac{24885308602613}{406425600}e_t^8
+\frac{6208604054057939}{40642560000}e_t^{10}
+\frac{1290387714679763381}{3901685760000}e_t^{12}
\Bigg)\nu
\Bigg] .
\end{align}
\end{widetext}
The relative-$1$PN term receives contributions from the $1$PN mass quadrupole, the Newtonian current quadrupole and mass octupole, the $1$PN orbital averaging measure, and the ADM mass correction. 

We note that the expressions above extend the results of~\cite{Bini:2025zvk} through $\mathcal{O}(e_t^{12})$, while reproducing the previously reported coefficients up to $\mathcal{O}(e_t^2)$. 
These results provide the harmonic-coordinate input required to determine the corresponding tail-of-tail contributions to the EOB Hamiltonian through $\mathcal{O}(e_t^{12})$ and fractional $1$PN order.

\section{Delaunay Averaged EOB Hamiltonian}
\label{sec:delEOB}
\subsection{The Matching Procedure}
The harmonic-coordinate results of the previous section can be transcribed into the
EOB framework by matching the Delaunay-averaged Hamiltonians~\cite{Bini:2025zvk},
\begin{equation}
\left\langle H_{\rm harmonic}^{\rm tt}\right\rangle
=
\left\langle H_{\rm EOB}^{\rm tt}\right\rangle,
\end{equation}
where
\begin{align}
\left\langle H_{\rm harmonic}^{\rm tt} \right\rangle
&\equiv
\frac{\displaystyle\oint H_{\rm harmonic}^{\rm tt}(t)\,dt}
{\displaystyle\oint dt},
\end{align}
and
\begin{align}
\left\langle H_{\rm EOB}^{\rm tt} \right\rangle
&\equiv
\frac{\displaystyle\oint H_{\rm EOB}^{\rm tt}(t)\,dt}
{\displaystyle\oint dt}.
\end{align}
Both sides should be expressed in terms of harmonic QK variables $(a_r,e_t)$, after applying the fractional-1PN map between the EOB and harmonic quasi-Keplerian parameters.
 
In the DJS gauge~\cite{Damour:2008qf}, the effective Hamiltonian reads
\begin{align}
\hat H_{\rm eff}
&=
\sqrt{
A(u,\nu)\bigl(1 + p_\phi^2 u^2 + p_r^2\,A(u,\nu)\bar D(u,\nu)+Q\bigr)
},
\end{align}
where $u = u_{\rm EOB} = 1/r_{\rm EOB}$, $p_r$ is the radial momentum, $p_\phi$ is the dimensionless canonical angular momentum, and $Q = Q(u,p_r;\nu)$ is a non-geodesic correction.

The connection to the real Hamiltonian follows from the EOB energy map~\cite{Buonanno:1998gg}
\begin{align}
H &= Mc^2\sqrt{1+2\nu(\hat H_{\rm eff}-1)}.
\end{align}
The radial potentials can be expanded as
\begin{align}
A(u,\nu) &= 1-2u+\sum_{n\geq3}a_n(\nu,\ln u)\,u^n,\\
\bar D(u,\nu) &= 1+\sum_{n\geq2}\bar d_n(\nu,\ln u)\,u^n,\\
Q(u,p_r;\nu) &= \sum_{m\geq2}p_r^{2m}\,q_{2m}(u,\nu).
\end{align}
\noindent
At the PN orders explored here, each tail-of-tail contribution contains a leading term entering the real Hamiltonian at $5.5$PN and a relative-$1$PN term entering at $6.5$PN.\footnote{The subscript on $a_n$, $\bar d_n$, and $q_n$ labels the power of $u$, not the PN order. Thus $a_{6.5}u^{13/2}$ and $\bar d_{5.5}u^{11/2}$ both contribute at 5.5PN, while $a_{7.5}u^{15/2}$ and $\bar d_{6.5}u^{13/2}$ both contribute at 6.5PN.} 
The corrections to the EOB metric potentials $A$ and $\bar{D}$ are given by
\begin{align}
\delta A &= a_{6.5}\,u^{13/2}+a_{7.5}\,u^{15/2},\\
\delta\bar D &= \bar d_{5.5}\,u^{11/2}+\bar d_{6.5}\,u^{13/2},
\end{align}
where the coefficients $\lbrace a_{6.5}, a_{7.5}, \bar{d}_{5.5}, \bar{d}_{6.5} \rbrace$ were determined from the tail-of-tail Hamiltonian in~\cite{Bini:2025zvk}. 
In contrast, the non-geodesic term admits the expansion
\begin{equation}
Q (u,p_r;\nu)=\sum_{m\ge2}\sum_{n} q_{2m,n}(\nu)\,p_r^{2m}u^{n},
\label{eq:Qdef}
\end{equation}
where the first subscript labels the power of the radial momentum and the second the power of $u$.
Through relative 1PN order, the tail-of-tail contribution in each $p_r^{2m}$ sector contains exactly two powers,
\begin{equation}
n\in\left\{\tfrac{13}{2}-m,\ \tfrac{15}{2}-m\right\}.
\label{eq:qstruct}
\end{equation}
For bound motion, $p_r^2 \sim \mathcal{O}(u)$, and hence $p_r^{2m}u^{13/2-m}\sim \mathcal{O}(u^{13/2})$ and $p_r^{2m}u^{15/2-m} \sim \mathcal{O}(u^{15/2})$. 
Since the Newtonian contribution to the Hamiltonian scales as $\mathcal{O}(u)$, these two terms enter the real Hamiltonian at $5.5$PN and $6.5$PN, respectively, independently of $m$.

Along the 1PN QK orbit, the radial momentum is $\mathcal{O}(e_t)$, so a term proportional to $p_r^{2m}$ first contributes to the orbital average at $\mathcal{O}(e_t^{2m})$.
The eccentricity expansion therefore turns our matching procedure into an iterative process for determining the EOB potentials.
The circular sector fixes the $A$-potential coefficients $a_{6.5}$ and $a_{7.5}$, while the $\mathcal{O}(e_t^2)$ sector fixes the $\bar D$-potential coefficients $\bar d_{5.5}$ and $\bar d_{6.5}$~\cite{Bini:2025zvk}.
Since the non-geodesic potential $Q$ begins at $p_r^4$ in the DJS gauge~\cite{Damour:2008qf}, it first enters at $\mathcal{O}(e_t^4)$.
Extending the expansion through $\mathcal{O}(e_t^{12})$ therefore allows us to determine the tail-of-tail contributions to the $Q$-potential coefficients up to $\mathcal{O}(p_r^{12})$.
Finally, we note that the truncation at $\mathcal{O}(e_t^{12})$ adopted here is purely computational. 
The procedure outlined here, following~\cite{Bini:2025zvk}, can be iterated to arbitrary order in $e_t$. 

At $1$PN, we can write the conserved energy $H^{(0)} = E$ and angular momentum $p^{(0)}_{\phi} = L$ as~\cite{Bini:2025zvk}
\begin{align}
E &= - \frac{1}{2 a^{{\rm e}}_r} + \eta^2 \frac{3 - \nu}{8} \frac{1}{a_r^{{{\rm e}}2}}, \\
L &=
\sqrt{a_r^{\rm e} \left(1 - e_r^{{\rm e} 2} \right)} + \eta^2 \frac{\frac{3}{2}+\frac{1}{2} e_r^{{\rm e} 2}}{\sqrt{a_r^{\rm e} \left(1 - e_r^{{\rm e} 2} \right) }} .
\end{align}
where $(a_r^{{\rm e}}, e_t^{{\rm e}}, e_r^{{\rm e}})$ are the orbital parameters in EOB coordinates. 
Following~\cite{Bini:2025zvk}, we can use energy conservation to determine $p_r^{(0)}$, finding
\begin{align}
p_r^{(0)} &=p_r^{\rm N} \left[ 1+\eta^2\left( \frac{1}{r}-\frac{1}{2a_r^{\rm e}} \right) \right].
\end{align}
Note that there is a typo in Ref.~\cite{Bini:2025zvk}, and our expressions differ from~\cite{Bini:2025zvk}, who have an additional factor of $r$ multiplying $1 / (2 a^e_r)$. 
The expression above follows directly from the $1$PN EOB energy-conservation equation and is the form adopted here.
The Newtonian prefactor is given by
\begin{align}
p_r^{\rm N} &= \sqrt{\frac{2}{r} -\frac{r^2 + a_r^{{\rm e}2} \left(1 - e_r^{{\rm e}2} \right)} {r^2 a_r^{\rm e}} }.
\end{align}
In order to relate our EOB Hamiltonian to the harmonic coordinate expressions in Eq.~\eqref{eq:HNtotharmonic} and Eq.~\eqref{eq:H1PNtotharmonic}, we can make use of the following transformations up to $1$PN~\cite{Damour:1981bh,Damour:1985der,Arun:2007rg,Bini:2025zvk}
\begin{align}
e_r^{\rm e} &= e^{\rm e}_t + \eta^2 \frac{3}{a_r^{\rm e}} e^{\rm e}_t, \\
a_r^{\rm e} &= a_r + \eta^2, \\
e^{\rm e}_t &= e_t - \eta^2 e_t \frac{\nu}{a_r}.
\end{align}

\subsection{The Tail-of-Tail Contributions to the EOB Potentials}
Matching the Delaunay averaged Hamiltonians order-by-order in $e_t$, we find the following $5.5$PN and $6.5$PN tail-of-tail corrections to the EOB metric potentials, in agreement with~\cite{Bini:2025zvk}
\begin{align}
a_{6.5}       &= \frac{13696}{525}\,\nu\pi, \\
a_{7.5}       &= -\nu\pi\left(\frac{512501}{3675} + \frac{10052}{225}\,\nu\right), \\
\bar{d}_{5.5} &= \frac{264932}{1575}\,\nu\pi, \\
\bar{d}_{6.5} &= -\nu\pi\left(\frac{21288791}{17640} + \frac{893149}{2450}\,\nu\right).
\end{align}
\noindent
Iterating the matching procedure up to $\mathcal{O}(e_t^{12})$, we find the following tail-of-tail corrections to the $Q$ potential:
\begin{align}
q_{4,4.5}  &= \frac{88703}{1890}\,\nu\pi, \\
q_{4,5.5} &= -\nu\pi\left(\frac{714117331}{846720} + \frac{224134961}{1058400}\,\nu\right), \\[0.5ex]
q_{6,3.5}  &= -\frac{2723471}{756000}\,\nu\pi, \\
q_{6,4.5}  &= \nu\pi\left(\frac{1783458013}{56448000} - \frac{1269199661}{127008000}\,\nu\right), \\[0.5ex]
q_{8,2.5}  &= \frac{5994461}{12700800}\,\nu\pi, \\
q_{8,3.5}  &= \nu\pi\left(\frac{12986592749}{22759833600} - \frac{11266879201}{28449792000}\,\nu\right), \\[0.5ex]
q_{10,1.5} &= -\frac{1657880149}{26127360000}\,\nu\pi, \\
q_{10,2.5} &= \nu\pi\left(\frac{6030897501941}{4551966720000} - \frac{1324939288373}{10241925120000}\,\nu\right), \\[0.5ex]
q_{12,0.5} &= \frac{23800130803}{3218890752000}\,\nu\pi, \\
q_{12,1.5} &= \nu\pi\Big(-\frac{1062624721887527}{2403438428160000} \nonumber \\ & \qquad \qquad \qquad \qquad + \frac{10982826421327}{163870801920000}\,\nu\Big).
\end{align}
As discussed earlier, for each $m$, the terms with $n=(13/2)-m$ and $n=(15/2)-m$ refer to the $5.5$PN and $6.5$PN sectors respectively, see the discussion around Eq.~\eqref{eq:Qdef} and Eq.~\eqref{eq:qstruct}.
The $\mathcal{O}(\nu)$ coefficients through $\mathcal{O}(p^8_r)$ are provided in Refs.~\cite{Bini:2016qtx,Bini:2020wpo}. 
Our calculation reproduces these terms exactly, and extends the calculation through $p^{12}_r$. 
The displayed $\mathcal{O}(\nu^2)$ terms provide qualitatively new $2$SF predictions for the non-geodesic potential, extending the 2SF predictions for $a_{7.5}$ and $\bar d_{6.5}$ originally derived in~\cite{Bini:2016qtx,Bini:2020wpo,Bini:2025zvk}. 

\section{Cross-Validation with Spectral Representations}
\label{sec:spectral}
The results of the preceding section were obtained by jointly expanding the multipole bilinears as a PN and eccentricity series before performing the Delaunay average. 
In this section, we follow a complementary route to the same result. 
We first decompose the source multipoles into their discrete frequency spectra, and then evaluate the hereditary kernel mode-by-mode. 
Expanding the resulting spectral sums through $\mathcal{O}(e_t^{12})$ provides a validation of the harmonic-coordinate calculation. 
This is very closely related to the spectral methods used in~\cite{Arun:2007rg,Arun:2007sg,Arun:2009mc,Damour:2015isa,Loutrel:2016cdw,Bini:2020wpo,Trestini:2025yyc,Liu:2026ovv}, as the problems are governed by the same source multipoles and orbital frequency spectra. 
This approach is independent at the level of the hereditary integral and orbital averaging, while using the same source multipoles and QK orbital dynamics, providing a useful cross-validation and self-consistency check of our results. 

The spectral representation also highlights two structural features of the calculation. 
First, the hereditary kernel acts diagonally in frequency space, multiplying each Fourier mode by $i\pi\,\mathrm{sgn}(\omega)$~\cite{Blanchet:1993ec,Damour:2015isa}. 
The Delaunay average then eliminates all cross-harmonics of the bilinears, collapsing them to weighted sums of squared Fourier amplitudes. 
Each averaged hereditary bilinear is thereby given by the negative of a positive-definite spectral norm, fixing its overall sign term by term~\cite{Arun:2007rg,Damour:2015isa}.


Second, combining the factor $\operatorname{sgn}(\omega)$ with the $2q+1$ time derivatives leads to harmonic weights proportional to $|p|^{2q+1}$. 
For even spectral weights, Parseval's theorem allows us to relate the harmonic sum to the orbital average of a local-in-time quantity. 
In particular, we find that $\sum_p p^{2k}|X_L(p)|^2$ is proportional to $\langle (X_L^{(k)})^2\rangle$. 
Within the QK parametrisation, such averages reduce to closed algebraic functions of the eccentricity, such as the Peters--Mathews enhancement function
$f(e_t)$~\cite{Peters:1963ux}.
For an odd weight, however, the identity $|p|^{2q+1}=p^{2q+1}\operatorname{sgn}(p)$ leaves an irreducible factor of $\operatorname{sgn}(p)$ or, equivalently, $\operatorname{sgn}(\omega)$. 
In the time-domian, multiplication by $\operatorname{sgn}(\omega)$ corresponds to a non-local principal-value kernel, see Eq.~\eqref{eq:PVFourier} below. 
It therefore cannot be generated by any finite combination of time derivatives. 
Consequently, the averaged hereditary bilinears cannot in general be expressed as orbital averages of local functionals of the motion. 
Instead, their corresponding eccentricity enhancement functions are naturally represented by infinite harmonic sums, for which simple closed-form expressions are in general not available, as seen in earlier work~\cite{Blanchet:1993ec,Arun:2007rg,Arun:2007sg,Loutrel:2016cdw}.

However, we note that Ref.~\cite{Loutrel:2016cdw} introduced an elegant resummation scheme for Kapteyn series~\cite{Kapteyn:1893,Watson:1944}, in which the Bessel functions are replaced by their uniform large-$p$ asymptotic expansions and the harmonic sums by integrals, yielding closed-form superasymptotic and hyperasymptotic approximants across the full range of eccentricities.  
Analogous resummations of the averaged tail-of-tail Hamiltonian could be explored, which we leave to future work.

At Newtonian order, a multipole $X_L$ is periodic in the mean anomaly and admits a Fourier expansion of the form~\cite{Peters:1963ux,Arun:2007rg}
\begin{align}
X_L(t)=\sum_{p \in \mathbb{Z}} X_L(p) e^{ip\ell}.
\label{eq:NewtonianFourier}
\end{align}
At $1$PN, however, we can no longer ignore the periastron advance and the orbital motion becomes quasiperiodic in the mean anomaly. 
Writing the multipole as a sum over azimuthal modes $m$, the periastron advance shifts the $p$th Fourier harmonic such that~\cite{Arun:2007rg}
\begin{align}
X_L(t) &=\sum_m\sum_{p\in\mathbb{Z}} X_{L,m}(p)e^{i(p+mk)\ell},
\label{eq:1PNFourier}
\end{align}
where the corresponding mode frequency is $\omega_{pm} = n (p + m k)$. 
In the following, we use these representations to construct the Newtonian and fractional-$1$PN spectral norms and compare their eccentricity expansions directly with the Delaunay-averaged results of the preceding section.

\subsection{Fourier--Bessel Amplitudes}
Let $F(\ell)$ be a $2\pi$-periodic function of the mean anomaly. 
Its Fourier expansion can be written as
\begin{align}
F(\ell)
&=\sum_{p=-\infty}^{\infty}
\mathcal{A}_p[F]\,e^{ip\ell},
\end{align}
where 
\begin{align}
\mathcal{A}_p[F]
&\equiv
\frac{1}{2\pi}\int_0^{2\pi}d\ell\,
F(\ell)e^{-ip\ell}
\nonumber\\
&=
\frac{1}{2\pi}\int_0^{2\pi}du\,
\rho(u)F(u)e^{-ip[u-e_t\sin u]}.
\label{eq:FBamp}
\end{align}
In the second line, we used Eq.~\eqref{eq:Kepler} together with the Jacobian
\begin{align}
\rho(u)\equiv\frac{d\ell}{du}=1-e_t\cos u,
\end{align}
to express the amplitudes in terms of the eccentric anomaly $u$. 
For the orbital functions encountered below, the weighted combination $\rho(u)F(u)$ reduces to a trigonometric polynomial in the eccentric anomaly. 
If we let $z=e^{iu}$, we see that these polynomials admit a finite Laurent expansion of the form
\begin{align}
\rho(u)F(u)=\sum_{j=-j_{\rm max}}^{j_{\rm max}} c_j z^j .
\end{align}
We can use the Laurent expansion to simplify Eq.~\eqref{eq:FBamp} by making use of the Jacobi-Anger identity~\cite{Abramowitz:1964}, which gives us a compact expression for the Fourier amplitudes in terms of the coefficients $c_j$ of the Laurent expansion
\begin{align}
\mathcal{A}_p[F] &= \sum_{j=-j_{\rm max}}^{j_{\rm max}} c_j J_{p-j}(p e_t).
\label{eq:FBmap}
\end{align}
Amplitudes of this form are closely related to the Hansen coefficients $X_p^{n,m}(e_t)$ familiar from celestial mechanics~\cite{Hansen:1855,Mikoczi:2015ewa}.
Although each Fourier amplitude is an exact finite combination of Bessel functions, we shall see that the eccentricity enhancement functions introduced below involve weighted sums of $|\mathcal{A}_p[F]|^2$ over infinitely many harmonics. 
Because the order and argument of the Bessel functions both grow with $p$, each such sum is a finite linear combination of Kapteyn series of the second kind~\cite{Kapteyn:1893,Watson:1944}, which converge for $0\le e_t<1$.

At Newtonian order, where $e_\phi=e_t$, we will find it useful to introduce
\begin{align}
\zeta_\pm(u) &\equiv \rho(u)e^{\pm iv(u)}  = \cos u-e_t\pm is\sin u,
\label{eq:zeta}
\end{align}
where $s\equiv\sqrt{1-e_t^2}$, so that $\zeta_-=\zeta_+^*$.
At $1$PN order, the periastron advance is treated separately through the shift to the mode frequencies $\omega_{pm}$.
We will denote the radial unit vector by
\begin{align}
\lambda_i\equiv\frac{r_i}{r},
\end{align}
such that at Newtonian order $\boldsymbol{\lambda}=(\cos v,\sin v,0)$ and $r_i=a_r\rho\lambda_i$.  
At $1$PN order, the separation direction will be given by $\boldsymbol{\lambda}=(\cos\phi,\sin\phi,0)$, with $\phi=(1+k)v$.

\subsection{The Hereditary Kernel as a Spectral Norm}
With an expression for the Fourier amplitudes at hand, the hereditary bilinears can be evaluated mode-by-mode by inserting the Fourier expansion directly into both the time derivatives of the MPM multipoles and the principal-value kernel introduced in Eq.~\eqref{eq:HilbertPV}. 
At Newtonian order, the $q$th time derivative of a multipole is trivially given by
\begin{align}
X_L^{(q)}(t) &=\sum_{p\in\mathbb{Z}} (ipn)^q X_L(p)e^{ip\ell},
\end{align}
and we can make use of the standard principal-value identity~\cite{Damour:2015isa}
\begin{align}
\operatorname{PV}\!\int_{-\infty}^{\infty}
\frac{dt'}{t'-t}e^{i\omega t'}
&=i\pi\operatorname{sgn}(\omega)e^{i\omega t} .
\label{eq:PVFourier}
\end{align}
As discussed above, the time derivatives and the hereditary kernel act diagonally on each Fourier mode, simplifying the resulting calculations.

Substituting these expressions into Eq.~\eqref{eq:HilbertPV} results in a double sum over the harmonics $p$ and $p'$. 
When we take the Delaunay average, only the terms satisfying $p'=-p$ will survive and the reality of the multipoles enforces $X_L(-p)=X_L^*(p)$. 
This allows us to derive an equivalent spectral representation of the averaged bilinears, analogous to the result derived in Sec.~\ref{sec:tailoftails}. 
Structurally, we find that the bilinears have the form~\cite{Arun:2007rg,Damour:2015isa}
\begin{align}
\left\langle H\!\left[X_L^{(q,q+1)}\right]\right\rangle
&=-\pi n^{2q+1}
\sum_{p\ne0}|p|^{2q+1}|X_L(p)|^2
\nonumber\\
&=-2\pi n^{2q+1}
\sum_{p\ge1}p^{2q+1}|X_L(p)|^2.
\label{eq:universalSpectralNorm}
\end{align}
Immediately, we can see that the averaged hereditary bilinears are the negative of a positive-definite spectral norm.
This provides a convenient check on the overall sign, and allows us to validate the sign of the mass-octupole contribution in Sec.~\ref{sec:tailoftails}.

\subsection{Newtonian Mass Quadrupole}
To illustrate this machinery, we derive the Newtonian mass quadrupole. 
Starting from Eq.~\eqref{eq:1PN_multipoles}, we have
\begin{align}
I_{ij}^{\rm N} &=\mu r_{\langle i}r_{j\rangle} = \mu a_r^2\rho^2\lambda_{\langle i}\lambda_{j\rangle},
\end{align}
though we will find it more convenient to work with the dimensionless reduced quadrupole moment~\cite{Damour:2015isa},
\begin{align}
\widehat I_{ij}
\equiv\frac{I_{ij}^{\rm N}}{\mu a_r^2}
=\rho^2\lambda_{\langle i}\lambda_{j\rangle}.
\end{align}

When constructing the bilinear in Eq.~\eqref{eq:I2bilinear}, we end up with a contribution of the form
\begin{align}
\lambda_{\langle i}\lambda_{j\rangle}
\lambda'_{\langle i}\lambda'_{j\rangle}
&=\frac16+\frac12\cos\!\left[2(v-v')\right].
\end{align}
Using $\zeta_\pm=\rho e^{\pm iv}$, we see that the quadrupole term can be expressed as
\begin{align}
\widehat I_{ij}(u)\widehat I_{ij}(u')
&=\frac16\rho^2(u)\rho^2(u')
+\frac14\zeta_+^2(u)\zeta_-^2(u')
\nonumber\\
&\quad+\frac14\zeta_-^2(u)\zeta_+^2(u'),
\label{eq:I2channels}
\end{align}
where the first term corresponds to a $m=0$ mode and the next two terms to the $m=\pm2$ modes. 

The problem therefore reduces to finding expressions for the Fourier amplitudes governing these modes, notably
\begin{align}
D_p &= \mathcal{A}_p[\rho^2],
&
C_p &= \mathcal{A}_p[\zeta_+^2],
\end{align}
with the $\zeta_-^2$ amplitude trivially following from $C_p$.
For $p \ne 0$, the amplitudes can be expressed in terms of Bessel functions, see App.~\ref{app:NewtonianMassQuadrupole},
\begin{align}
\label{eq:Dp}
D_p&=-\frac{2}{p^2} J_p (p e_t), \\
C_p&=-\frac{2(1+s^2-2p s^3)}{e_t^2p^2} J_p (p e_t)
-\frac{4s(1-p s)}{e_t p^2} J'_p (p e_t),
\label{eq:Cp}
\end{align}
where a prime denotes differentiation with respect to the argument and the circular limit is understood by continuity. 
We note that the zero-frequency modes do not contribute as the hereditary bilinear contains time derivatives. 
The spectral norm for the mass quadrupole bilinear can therefore be written as
\begin{align}
\widehat I_{ij}(p)\widehat I_{ij}^*(p)
&=\frac16|D_p|^2
+\frac14|C_p|^2
+\frac14|C_{-p}|^2.
\label{eq:I2-channel-norm}
\end{align}
Evaluating the expression in terms of the Bessel functions, we find that this is directly related to the Peters--Mathews enhancement function~\cite{Peters:1963ux}, such that
\begin{align}
g(p,e_t) &= \frac{p^6}{16} \widehat I_{ij}(p)\widehat I_{ij}^*(p),
\end{align}
normalised such that $g(2,0)=1$.
Using Eq.~\eqref{eq:universalSpectralNorm}, with $q=3$, we find~\cite{Bini:2020wpo}
\begin{align}
2\left\langle H\!\left[I_{2,\eta^0}^{(3,4)}\right]\right\rangle
&=-\frac{\pi\nu^2}{a_r^{13/2}}\mathcal{I}_0(e_t),
\label{eq:I0spectral}
\end{align}
where
\begin{align}
\mathcal{I}_0(e_t)
&= 4 \, \sum_{p\ge1} \, p^7 \widehat I_{ij}(p) \widehat I_{ij}^*(p)
\nonumber\\
&=128 \, \sum_{p\ge1} \, \frac{p}{2} g(p,e_t)
\nonumber\\
&=128 \, \varphi(e_t).
\label{eq:I0def}
\end{align}
where $\varphi(e_t)$ is the dimensionless eccentricity factor introduced in~\cite{Arun:2007rg}, itself related to the earlier factor introduced in~\cite{Blanchet:1993ec} by a factor of the Peters-Mathews enhancement function $f(e_t)$~\cite{Peters:1963ux}. 

The above result is in full agreement with the earlier calculation presented in Refs.~\cite{Damour:2015isa,Bini:2020wpo}. 
These are closely related to the expressions obtained in~\cite{Arun:2007rg} for the first-order tail contribution to the averaged gravitational-wave flux and to the expressions in~\cite{Damour:2015isa} for the averaged second-order tails. 
This should come as no surprise, since the calculations involve the same underlying source multipoles but with different harmonic weights.
Eccentricity expanding the enhancement function in Eq.~\eqref{eq:I0spectral} gives
\begin{align}
\mathcal{I}_0(e_t)
&=128+\frac{4670}{3}e_t^2+\frac{42955}{6}e_t^4
\nonumber\\
&\quad+\frac{6204647}{288}e_t^6
+\frac{352891481}{6912}e_t^8
\nonumber\\
&\quad+\frac{286907786543}{2764800}e_t^{10}
\nonumber\\
&\quad
+\frac{6287456255443}{33177600}e_t^{12}
+\mathcal{O}(e_t^{14}),
\label{eq:I0series}
\end{align}
in complete agreement with the results obtained in Eq.~\eqref{eq:HI2N}.

\subsection{Current Quadrupole and Mass Octupole}
The Newtonian current quadrupole and mass octupole both carry the mass-difference factor, 
\begin{align}
\delta = \sqrt{1-4\nu}.
\end{align}
We opt to explicitly factor this out of the corresponding enhancement functions by defining the following reduced moments
\begin{align}
I_{ijk}^{\rm N}
&\equiv\mu\delta a_r^3\,\overline I_{ijk},
&
\overline I_{ijk}
&=\rho^3\lambda_{\langle i}\lambda_j\lambda_{k\rangle},
\\
J_{ij}^{\rm N}
&\equiv\mu\delta a_r^3n\,\overline J_{ij},
&
\overline J_{ij}
&=s\rho\,\lambda_{\langle i}\widehat L_{j\rangle},
\end{align}
where $\widehat L_i$ denotes the unit vector normal to the orbital plane. 
From Eq.~\eqref{eq:universalSpectralNorm}, the current quadrupole and mass octupole contributions are given by
\begin{align}
\label{eq:J2spectral}
2\left\langle H\!\left[J_2^{(3,4)}\right]\right\rangle
&=-(1-4\nu)\frac{\pi\nu^2}{a_r^{15/2}}
\mathcal J_0(e_t),
\\
2\left\langle H\!\left[I_3^{(4,5)}\right]\right\rangle
&=-(1-4\nu)\frac{\pi\nu^2}{a_r^{15/2}}
\mathcal O_0(e_t),
\label{eq:I3spectral}
\end{align}
where
\begin{align}
\mathcal J_0(e_t)
&\equiv4\sum_{p\geq1}p^7
\left|\overline J_{ij}(p)\right|^2
=\gamma(e_t),
\\
\mathcal O_0(e_t)
&\equiv4\sum_{p\geq1}p^9
\left|\overline I_{ijk}(p)\right|^2
=\frac{49209}{5}\beta(e_t).
\end{align}
The enhancement functions $\gamma(e_t)$ and $\beta(e_t)$ are exactly those derived in~\cite{Arun:2007rg}. 
In particular, $\gamma(e_t)$ is given by
\begin{align}
\gamma(e_t)
&=2s^2\sum_{p=1}^{\infty}p^5
\left[
J_p'^2(pe_t)
+\frac{s^2}{e_t^2}J_p^2(pe_t)
\right],
\label{eq:gamma}
\end{align}
and $\beta(e_t)$ by
\begin{align}
\label{eq:beta-bessel}
\beta(e_t)
&=\frac{1}{49209}
\sum_{p=1}^{\infty}p^9
\Bigg[
\frac52\left(|U_p^+|^2+|U_p^-|^2\right) \\ 
&\nonumber \qquad \qquad \qquad 
+\frac32\left(|V_p^+|^2+|V_p^-|^2\right)
\Bigg],
\end{align}
where we have introduced the Fourier amplitudes
\begin{align}
\label{eq:UpVp}
U_p^\pm&\equiv\mathcal A_p[\zeta_\pm^3],
&
V_p^\pm&\equiv\mathcal A_p[\rho^2\zeta_\pm],
\end{align}
whose explicit forms are given in App.~\ref{app:bessel-amplitudes}. 
The resulting eccentricity expansions of $\mathcal{J}_0(e_t)$ and $\mathcal{O}_0(e_t)$ can be shown to reduce to
\begin{align}
\mathcal{J}_0(e_t)
&=1+30e_t^2+\frac{7321}{32}e_t^4
\nonumber\\
&\quad+\frac{279485}{288}e_t^6
+\frac{73162261}{24576}e_t^8
\nonumber\\
&\quad+\frac{3426964009}{460800}e_t^{10}
\nonumber\\
&\quad
+\frac{105664935873}{6553600}e_t^{12}
+\mathcal{O}(e_t^{14}),
\label{eq:J0series}\\
\mathcal{O}_0(e_t)
&=\frac{49209}{5}+\frac{910188}{5}e_t^2
+\frac{367780251}{320}e_t^4
\nonumber\\
&\quad+\frac{849097483}{192}e_t^6
+\frac{104770539213}{8192}e_t^8
\nonumber\\
&\quad+\frac{47178621858259}{1536000}e_t^{10}
\nonumber\\
&\quad+\frac{25457292864087389}{393216000}e_t^{12}
+\mathcal{O}(e_t^{14}).
\label{eq:O0series}
\end{align}
Inserting these expansions into Eq.~\eqref{eq:J2spectral} and Eq.~\eqref{eq:I3spectral} respectively, we find complete agreement with the earlier results derived in Eq.~\eqref{eq:HavJ2} and Eq.~\eqref{eq:HavI3}.

\subsection{Fractional-$1$PN Mass Quadrupole}
At fractional $1$PN order, the quadrupole amplitudes and their spectral frequencies receive relativistic corrections due to the periastron advance.
The $1$PN contribution to the Delaunay averaged Hamiltonian can be written as
\begin{align}
\label{eq:I21PNspectral}
2\left\langle H \! \left[ I_{2,\eta^2}^{(3,4)} \right] \right\rangle &= \frac{\pi\nu^2}{a_r^{15/2}} \mathcal I_1(e_t,\nu),
\end{align}
where we have introduced the $1$PN enhancement function
\begin{align}
    \mathcal I_1(e_t,\nu) &=A_{I_2}(e_t)+\nu B_{I_2}(e_t).
\end{align}
In the hereditary-flux notation of Refs.~\cite{Arun:2007rg,Arun:2007sg}, these functions are related by
\begin{align}
A_{I_2}(e_t)
&=\frac{13696}{21}\alpha(e_t)
+\left(2496-\frac{1664}{1-e_t^2}\right)\varphi(e_t),
\\
B_{I_2}(e_t)
&=-\frac{5696}{21}\theta(e_t)
-\frac{1024}{3}\varphi(e_t),
\end{align}
where $\alpha(e_t)$ and $\theta(e_t)$ are also standard hereditary enhancement functions introduced in Refs.~\cite{Arun:2007rg,Arun:2007sg}. 
Eccentricity expanding these functions, we find
\begin{align}
A_{I_2}(e_t)
&=\frac{31168}{21}
+\frac{256805}{21}e_t^2
+\frac{339379}{12}e_t^4
-\frac{309479}{1728}e_t^6
\nonumber\\
&\quad
-\frac{19714929593}{96768}e_t^8
-\frac{1527310235551}{1843200}e_t^{10}
\nonumber\\
&\quad
-\frac{351712023425809}{154828800}e_t^{12}
+\mathcal{O}(e_t^{14}),
\label{eq:AI2-series}
\\[1ex]
B_{I_2}(e_t)
&=-\frac{4288}{7}
-\frac{158993}{21}e_t^2
-\frac{146445}{4}e_t^4
-\frac{67251767}{576}e_t^6
\nonumber\\
&\quad
-\frac{28402847971}{96768}e_t^8
-\frac{388346663887}{614400}e_t^{10}
\nonumber\\
&\quad
-\frac{566906059048153}{464486400}e_t^{12}
+\mathcal{O}(e_t^{14}).
\label{eq:BI2-series}
\end{align}
When inserted into Eq.~\eqref{eq:I21PNspectral}, we find complete agreement with the result derived in Eq.~\eqref{eq:HI21PN}.

\section{Redshift Invariants}
\label{sec:redshift}
The first law of binary mechanics provides a link between the conservative tail-of-tail dynamics derived above and gauge-invariant observables~\cite{LeTiec:2011ab,Barausse:2011dq,LeTiec:2015kgg,
Bini:2016qtx,Bini:2016cje,Bini:2019nra,Bini:2020wpo}, such as the Detweiler--Barack--Sago
redshift~\cite{Detweiler:2008ft,Barack:2009ux}. 
It is naturally formulated as a variational statement about the ADM mass $\mathcal{M}$ of the system~\cite{LeTiec:2011ab,LeTiec:2015kgg}, which can be related to the binding energy by $\mathcal{M}=M+E_{\rm b}$.
Comparison with the redshift invariants provides both a direct test of the small-mass-ratio limit of the averaged Hamiltonian and a natural point of comparison with second-order self-force predictions.
Since tail effects are inherently hereditary, we formulate this comparison in terms of the localised Delaunay-averaged tail-of-tail Hamiltonian $\langle H_{\rm tt}\rangle$, for which the first law retains its standard form~\cite{Blanchet:2017rcn}.

For a non-spinning binary on a stable bound orbit, the conservative motion may be parametrised by the radial and azimuthal actions,
\begin{equation}
J_r=\frac{1}{2\pi}\oint p_r\,dr,
\qquad
J_\phi=L=p_\phi .
\end{equation}
The fundamental frequencies associated to the angle variables are given by
\begin{equation}
\Omega_r=\frac{2\pi}{T_r},
\qquad
\Omega_\phi=\frac{\Phi}{T_r},
\end{equation}
where $T_r$ denotes the radial period and $\Phi$ the azimuthal phase accumulated during one radial cycle. 
The averaged Detweiler-Barack-Sago redshift of body $a$ is then defined by~\cite{Detweiler:2008ft,Sago:2008id,Blanchet:2009sd,
Blanchet:2010zd,Barack:2011ed}
\begin{equation}
\langle z_a\rangle
 =\frac{1}{T_r}\int_0^{T_r}\frac{d\tau_a}{dt}\,dt
 =\frac{\mathcal{T}_a}{T_r},
\end{equation}
where $\mathcal{T}_a$ is the proper-time duration of one radial cycle.
We also introduce the corresponding inverse redshift invariant,
\begin{equation}
U_a\equiv\frac{1}{\langle z_a\rangle}
=\frac{T_r}{\mathcal{T}_a}.
\end{equation}

To make the connection to GSF results, we need to work in the small-mass-ratio limit. 
Here, we adopt the conventions of Ref.~\cite{Bini:2016cje}, namely
\begin{equation}
m_1\ll m_2, \qquad q=\frac{m_1}{m_2}, \qquad M=m_2(1+q),
\end{equation}
where the larger body sources the background Schwarzschild geometry.
For convenience, we will denote the averaged redshift of the smaller body and its inverse by
\begin{equation}
z\equiv\langle z_1\rangle, \qquad U\equiv\frac{1}{z}.
\end{equation}

For a generic stable bound eccentric orbit, the averaged redshift may be parametrised by the mass ratio and the two dimensionless fundamental frequencies
\begin{equation}
\widehat\Omega_r\equiv m_2\Omega_r,
\qquad
\widehat\Omega_\phi\equiv m_2\Omega_\phi ,
\end{equation}
which are held fixed in the self-force expansions below.
For fixed frequencies, we can equivalently describe the system in terms of a reference Schwarzschild geodesic using the Darwin parameters $(p,e)$~\cite{Darwin:1959var,Barack:2011ed}. 
At fixed $m_2\Omega_r$ and $m_2\Omega_\phi$, the self-force expansion of $z$ and $U$ take the form
\begin{align}
z &= z_{\rm geo} +q\,\Delta z^{\rm 1SF} + q^2\,\Delta z^{\rm 2SF} +\mathcal{O}(q^3),
\\
U &= U_{\rm geo} + q\,\Delta U^{\rm 1SF} +q^2\,\Delta U^{\rm 2SF} +\mathcal{O}(q^3),
\end{align}
where $U_{\rm geo}=1/z_{\rm geo}$. 
Expanding the identity $U=1/z$ order by order in $q$ yields
\begin{align}
\Delta U^{\rm 1SF} &= -\frac{\Delta z^{\rm 1SF}}{z_{\rm geo}^2},
\label{eq:U-z-1SF}
\\
\Delta U^{\rm 2SF} &= -\frac{\Delta z^{\rm 2SF}}{z_{\rm geo}^2} +\frac{\bigl(\Delta z^{\rm 1SF}\bigr)^2}{z_{\rm geo}^3}.
\label{eq:U-z-2SF}
\end{align}
The remaining task is therefore to evaluate the tail-of-tail contribution to the fixed-frequency mass derivatives at each self-force order. 

The first law of binary black hole mechanics can be written in terms of the action variables as~\cite{LeTiec:2011ab,LeTiec:2015kgg}
\begin{equation}
\delta \mathcal{M} = \Omega_r\,\delta J_r + \Omega_\phi\,\delta L + \sum_a \langle z_a\rangle\,\delta m_a .
\label{eq:eccentric-first-law}
\end{equation}
This variational law relates neighboring binary configurations, connecting variations in the ADM mass, radial action, orbital angular momentum, and component masses through the orbital frequencies and averaged redshifts.
The analysis of Ref.~\cite{Blanchet:2017rcn} shows that the first law retains its standard form for conservative hereditary dynamics, provided one works with a localised Hamiltonian, such as the Delaunay averaged Hamiltonian derived here.

The first law relates the averaged redshift to the variation of the ADM mass with respect to the corresponding mass at fixed actions~\cite{LeTiec:2011ab,LeTiec:2015kgg,Blanchet:2012at,Blanchet:2013txa,Blanchet:2017rcn,Trestini:2025yyc}
\begin{equation}
\langle z_a\rangle
 = \left.\frac{\partial \mathcal{M}}{\partial m_a}\right|_{J_r,L,m_{b\ne a}} .
\label{eq:redshift-fixed-actions}
\end{equation}
However, the self-force expansions introduced above are naturally given in terms of fixed fundamental frequencies rather than fixed actions. 
Following~\cite{LeTiec:2015kgg}, we can replace the action variables by frequencies via the Legendre transform
\begin{equation}
\mathcal R(\Omega_r,\Omega_\phi;m_a) = \mathcal{M}-\Omega_r J_r-\Omega_\phi L ,
\label{eq:frequency-space-R}
\end{equation}
whose variation reads
\begin{equation}
\delta\mathcal R = -J_r\,\delta\Omega_r-L\,\delta\Omega_\phi + \sum_a\langle z_a\rangle\,\delta m_a.
\end{equation}
The redshift then follows as a mass derivative at fixed frequencies,
\begin{equation}
\langle z_a\rangle = \left.\frac{\partial\mathcal R}{\partial m_a} \right|_{\Omega_r,\Omega_\phi,m_{b\ne a}},
\label{eq:redshift-from-R}
\end{equation}
which is the form we adopt when comparing against GSF results.

Working to linear order in the tail-of-tail interaction, its contribution to $\mathcal{M}$, and hence to $\mathcal{R}$, is given by the Delaunay-averaged Hamiltonian evaluated at the zeroth-order action variables,
\begin{equation}
\label{eq:Rtt_to_Htt}
\mathcal R_{\rm tt}(\Omega_r,\Omega_\phi;m_a)
= \left\langle H_{\rm tt}\right\rangle\!
\left(J_i^{(0)}(\Omega_r,\Omega_\phi;m_a)\right).
\end{equation}

Following the conventions adopted in eccentric self-force calculations~\cite{Barack:2011ed,Hopper:2015icj,Bini:2016cje,Munna:2022gio}, we expand in powers of $q$ at fixed $m_2$-rescaled frequencies $(\widehat\Omega_r,\widehat\Omega_\phi)$.
Since $m_1=q\,m_2$, the mass derivative in Eq.~\eqref{eq:redshift-from-R} taken at fixed $m_2$ is equivalent to a derivative with respect to $q$, and the tail-of-tail contribution to the redshift can be written as
\begin{equation}
\Delta z_{\rm tt} =\left.\frac{\partial}{\partial q} \left(\frac{\mathcal R_{\rm tt}}{m_2}\right) \right|_{\widehat\Omega_r,\widehat\Omega_\phi,m_2}.
\label{eq:redshift-frequency-derivative}
\end{equation}
Expanding in the mass ratio, the tail-of-tail contributions to the redshift at each self-force order are then defined by
\begin{equation}
\Delta z_{\rm tt} = q\,\Delta z_{\rm tt}^{\rm 1SF} + q^2\,\Delta z_{\rm tt}^{\rm 2SF} +\mathcal O(q^3),
\label{eq:delta-z-tt-expansion}
\end{equation}
with the corresponding contributions to the inverse redshift, 
$\Delta U_{\rm tt}^{\rm 1SF}$ and $\Delta U_{\rm tt}^{\rm 2SF}$, following from Eqs.~\eqref{eq:U-z-1SF} and \eqref{eq:U-z-2SF}.

Finally, we note that one could adopt alternative conventions, e.g. Refs.~\cite{Barausse:2011dq,LeTiec:2011dp}.
There they choose to expand in powers of $\nu$ at fixed $M$-rescaled frequencies, which necessarily changes the interpretation of the expansion coefficients~\cite{Bini:2016cje}. 
Since $M\Omega=(1+q)\,m_2\Omega$ and $\nu=q/(1+q)^2$, we would expect the two sets of expansion coefficients to differ from first order in the mass ratio onwards. 

\subsection{First-Order Self-Force Comparison}
We begin with the first-order self-force sector, for which independent results are available in the literature, notably Ref.~\cite{Munna:2022gio}.
The averaged tail-of-tail Hamiltonian derived above can be written as
\begin{align}
\label{eq:Htt-q-expansion}
\frac{\langle H_{\rm tt}\rangle}{M}
={}&
\pi\nu^2
\left[
\frac{F_0(e_t)}{a_r^{13/2}}
-\frac{F_1(e_t)+\nu G_1(e_t)}{a_r^{15/2}}
\right]
\nonumber\\
&\quad
+\mathcal O(a_r^{-17/2},e_t^{14}),
\end{align}
where $F_0$ is the eccentricity polynomial appearing in the
Newtonian-order result in Eq.~\eqref{eq:HNtotharmonic}, while $F_1$
and $G_1$ are respectively the $\nu$-independent and linear-in-$\nu$
parts of the fractional-$1$PN result in
Eq.~\eqref{eq:H1PNtotharmonic}.

As $M=m_2(1+q)$, the tail-of-tail contribution to the
frequency-space potential is given by
\begin{equation}
\frac{\mathcal R_{\rm tt}}{m_2}
=(1+q)\frac{\langle H_{\rm tt}\rangle}{M}.
\label{eq:Rtt-normalisation}
\end{equation}
Since $\nu = q / {(1+q)^2}$,  the $q$-dependent prefactor satisfies
\begin{equation}
(1+q)\nu^2=q^2+\mathcal O(q^3).
\end{equation}
Moreover, the term proportional to $\nu G_1$ in Eq.~\eqref{eq:Htt-q-expansion} first contributes at order $q^3$, and can be ignored here. 
The terms relevant to $\mathcal R_{\rm tt}$ at 1SF order are therefore
\begin{equation}
\frac{\mathcal R_{\rm tt}}{m_2} = q^2\mathcal H_{\rm tt}^{[q^2]}(a_r,e_t) + \mathcal O(q^3),
\end{equation}
where
\begin{equation}
\mathcal H_{\rm tt}^{[q^2]}(a_r,e_t) = \pi\left[ \frac{F_0(e_t)}{a_r^{13/2}} -\frac{F_1(e_t)}{a_r^{15/2}} \right].
\label{eq:Htt-q2-coefficient}
\end{equation}

To implement the fixed-frequency derivative, we use the Darwin parameters $(p,e)$ to label the reference Schwarzschild geodesic with the prescribed dimensionless frequencies $(\widehat\Omega_r,\widehat\Omega_\phi)$~\cite{Darwin:1959var}\footnote{In the strong-field regime near the separatrix, the map between $(p,e)$ and $(\widehat\Omega_r,\widehat\Omega_\phi)$ can fail to be invertible with physically distinct geodesics sharing frequencies~\cite{Barack:2011ed,Warburton:2013yj}. We do not expect this to be a problem for the weak-field expansions considered here.}.
This follows the convention adopted in eccentric self-force calculations~\cite{Barack:2011ed,Munna:2022gio}.
Since the geodesic frequency map between $(p,e)$ and $(\widehat\Omega_r,\widehat\Omega_\phi)$ is independent of the mass ratio, derivatives taken at fixed $(p,e)$ coincide with the fixed-frequency derivative of Eq.~\eqref{eq:redshift-frequency-derivative}.
At fixed $(p,e)$, the QK orbital elements admit expansions of the form
\begin{align}
a_r &= a_{r,\rm geo} + \mathcal O(q),
\\
e_t &= e_{t,\rm geo} + \mathcal O(q).
\end{align}
Because $\mathcal R_{\rm tt}$ begins at order $q^2$, the $\mathcal O(q)$ shifts to the orbital parameters first affect the mass derivative at 2SF order.  
The 1SF result can therefore be directly evaluated along the reference Schwarzschild geodesic.

At the PN accuracy used here, the test-particle relations are given by~\cite{Cutler:1994pb,Barack:2011ed,
Akcay:2015pza,Hopper:2015icj,Forseth:2015oua,Skoupy:2024jsi}
\begin{align}
a_{r,\rm geo}
&=\frac{p}{1-e^2}-1+\mathcal O(p^{-1}),
\label{eq:ar-p-map}
\\
e_{t,\rm geo}
&=e\left[
1-3\frac{1-e^2}{p}
+\mathcal O(p^{-2})
\right],
\label{eq:et-e-map}
\\
z_{\rm geo}
&=1-\frac{3}{2}\frac{1-e^2}{p}
+\mathcal O(p^{-2}).
\label{eq:zgeo-1PN}
\end{align}

Taking the derivative in
Eq.~\eqref{eq:redshift-frequency-derivative} at fixed $(p,e)$ gives
\begin{equation}
\left.
\frac{\partial}{\partial q}
\left(\frac{\mathcal R_{\rm tt}}{m_2}\right)
\right|_{p,e}
=
2q\,
\mathcal H_{\rm tt}^{[q^2]}
(a_{r,\rm geo},e_{t,\rm geo})
+\mathcal O(q^2).
\end{equation}
Matching to Eq.~\eqref{eq:delta-z-tt-expansion} at order $q$, and using Eq.~\eqref{eq:U-z-1SF}, we find
\begin{align}
\label{eq:z1SF-from-H}
\Delta z_{\rm tt}^{\rm 1SF} &= 2\, \mathcal H_{\rm tt}^{[q^2]} (a_{r,\rm geo},e_{t,\rm geo}),
\\
\Delta U_{\rm tt}^{\rm 1SF} &= -\frac{ 2\,\mathcal H_{\rm tt}^{[q^2]} (a_{r,\rm geo},e_{t,\rm geo}) }{z_{\rm geo}^2}.
\label{eq:U1SF-from-H}
\end{align}
Substituting Eqs.~\eqref{eq:ar-p-map}--\eqref{eq:zgeo-1PN} into this expression, we obtain
\begin{align}
\Delta U_{\rm tt}^{\rm 1SF} ={}& \pi\left[ \frac{\mathcal U_{11/2}^{(1)}(e)}{p^{13/2}} +\frac{\mathcal U_{13/2}^{(1)}(e)}{p^{15/2}}
\right]
\nonumber\\
&\quad +\mathcal O(p^{-17/2},e^{14}),
\label{eq:U1SF-tail-result}
\end{align}
where the subscripts denote the corresponding PN orders and the eccentricity-expanded coefficients are given by
\begin{align}
\mathcal U_{11/2}^{(1)}(e)
={}&
-\frac{13696}{525}
-\frac{232618}{1575}e^2
+\frac{430889}{3150}e^4
\nonumber\\
&+\frac{18404963}{151200}e^6
-\frac{7527343}{145152}e^8
\nonumber\\
&-\frac{22605883901}{1451520000}e^{10}
\nonumber\\
&-\frac{104442581123}{17418240000}e^{12}
+\mathcal O(e^{14}),
\label{eq:U11half-result}
\\[1ex]
\mathcal U_{13/2}^{(1)}(e)
={}&
\frac{81077}{3675}
+\frac{2687231}{4410}e^2
+\frac{13695499}{47040}e^4
\nonumber\\
&-\frac{1953554329}{1411200}e^6
+\frac{112283726591}{812851200}e^8
\nonumber\\
&+\frac{7790443135159}{40642560000}e^{10}
\nonumber\\
&+\frac{398929659254227}{7803371520000}e^{12}
+\mathcal O(e^{14}).
\label{eq:U13half-result}
\end{align}
In the circular limit, these expressions reduce to the known half-integer redshift coefficients~\cite{Shah:2013uya,Bini:2013rfa,Blanchet:2013txa,Blanchet:2014bza}. 
More generally, the leading-order eccentricity function in Eq.~\eqref{eq:Htt-q-expansion} satisfies the exact relation
\begin{equation}
F_0(e)=\frac{6848}{525} \, \varphi(e),
\end{equation}
which follows from Eqs.~\eqref{eq:I0spectral} and \eqref{eq:I0def}, together with the normalisation of the Newtonian mass-quadrupole contribution to the tail-of-tail Hamiltonian. 
Using the leading-order parts of the geodesic relations in Eqs.~\eqref{eq:ar-p-map}--\eqref{eq:zgeo-1PN}, the complete $5.5$PN eccentricity dependence can therefore be written as
\begin{equation}
\mathcal U_{11/2}^{(1)}(e)
=-\frac{13696}{525}(1-e^2)^{13/2}\varphi(e),
\end{equation}
which agrees exactly, to all orders in eccentricity, with the independent 1SF result of Ref.~\cite{Munna:2022gio}. 
Eccentricity expanding this expression reproduces the coefficients displayed above, after accounting for the overall factor of $1/p$ in the PN-series convention of Ref.~\cite{Munna:2022gio}. 
At $6.5$PN order, the two results agree coefficient by coefficient through $\mathcal O(e^{12})$, again after accounting for the overall factor of $1/p$ in~\cite{Munna:2022gio}.
These comparisons provide non-trivial checks of both the leading and fractional-$1$PN eccentricity dependence of the averaged tail-of-tail Hamiltonian in the small-mass-ratio limit.

\subsection{Eccentric Second-Order Self-Force Prediction}
\label{sec:2SF-eccentric}
As discussed in Sec.~\ref{sec:redshift}, extending the calculation to $2$SF requires the variation of the orbital elements with the mass ratio at fixed dimensionless fundamental frequencies $(\widehat\Omega_r,\widehat\Omega_\phi)$~\cite{Gralla:2008fg,Gralla:2012db,Pound:2014koa,Pound:2015wva}.
This is a generic feature of second-order perturbation theory, and also arises in the two-timescale formulation of the self-force problem~\cite{Hinderer:2008dm,Miller:2020bft,Pound:2021qin}.
The results derived here provide weak-field benchmarks for the ongoing $2$SF program, which has delivered conservative invariants, fluxes, and waveforms for quasi-circular orbits on a Schwarzschild background~\cite{Pound:2019lzj,Warburton:2021kwk,Wardell:2021fyy,Upton:2025bja}, and for which extensions to eccentric orbits are under active development~\cite{Leather:2023dzj,Wei:2025lva}, supported by high-order analytic $1$SF results for eccentric orbits~\cite{Hopper:2015icj,Munna:2022gio,Munna:2023wce}.
In this context, $\mathcal{R}_{\rm tt}$ starts at $\mathcal O(q^2)$, so its $2$SF contribution requires, beyond the explicit mass-ratio dependence of $\langle H_{\rm tt}\rangle$, only the linear-in-$q$ variation of the orbital elements at fixed frequencies.

For convenience, we introduce the following variables,
\begin{equation}
x\equiv e^2,
\qquad
w\equiv1-e^2=1-x,
\qquad
x_t\equiv e_t^2,
\end{equation}
and express the eccentricity polynomials in Eq.~\eqref{eq:Htt-q-expansion} as
functions of $x_t$. 
We write the averaged tail-of-tail Hamiltonian as
\begin{align}
\frac{\langle H_{\rm tt}\rangle}{M}
&=\pi\nu^2\,\mathcal B(q;p,e),
\label{eq:Htt-def}
\end{align}
where
\begin{align}
\mathcal B(q;p,e)
&=\frac{F_0(x_t)}{a_r^{13/2}}
-\frac{F_1(x_t)+\nu G_1(x_t)}{a_r^{15/2}}.
\end{align}
Combining this definition with Eq.~\eqref{eq:Rtt-normalisation} gives
\begin{equation}
\frac{\mathcal R_{\rm tt}}{m_2}
=\pi(1+q)\nu^2\,\mathcal B(q;p,e).
\label{eq:Rtt-B}
\end{equation}
The mass-ratio dependence of $\mathcal B$ arises through $\nu(q)$ and the fixed-frequency orbital elements $a_r(q;p,e)$ and $x_t(q;p,e)$.

As before, we use $(p,e)$ to label the reference Schwarzschild geodesic associated with the fixed frequency pair.  
With this convention, the energy and angular momentum maps can be written as
\begin{align}
\varepsilon(q) &=\varepsilon_{\rm geo}+q\,\delta_q\varepsilon+\mathcal O(q^2),
\\
j(q) &=j_{\rm geo}+q\,\delta_qj+\mathcal O(q^2).
\end{align}
Differentiating the QK frequency map~\cite{Damour:1988mr,Memmesheimer:2004cv} at fixed $(\widehat\Omega_r,\widehat\Omega_\phi)$, following the approach of Refs.~\cite{Barack:2011ed,LeTiec:2015kgg}, yields
\begin{align}
\delta_q\varepsilon
&\equiv
\left.\frac{\partial\varepsilon}{\partial q}
\right|_{\widehat\Omega_r,\widehat\Omega_\phi,q=0}
=\frac{2}{3}\varepsilon_{\rm geo}
+\frac{3}{4}\varepsilon_{\rm geo}^2
+\mathcal O(\varepsilon_{\rm geo}^3),
\label{eq:epsilon-fixed-frequency-shift}
\\
\label{eq:j-fixed-frequency-shift}
\delta_qj
&\equiv
\left.\frac{\partial j}{\partial q}
\right|_{\widehat\Omega_r,\widehat\Omega_\phi,q=0}
=\frac{2}{3}j_{\rm geo}
+\frac{5}{12}(j_{\rm geo}-6)\varepsilon_{\rm geo} \\
\nonumber
&\qquad \qquad \qquad \qquad \qquad \qquad +\mathcal O(\varepsilon_{\rm geo}^2).
\end{align}
The $\mathcal O(\varepsilon_{\rm geo})$ term in Eq.~\eqref{eq:j-fixed-frequency-shift} requires the 2PN periastron advance, which was provided in Sec.~\ref{sec:qk}.

Combining these shifts with the QK equations of motion and the Schwarzschild
relations between $(\varepsilon,j)$ and $(p,e)$ gives~\cite{Darwin:1959var,Cutler:1994pb,Barack:2011ed}
\begin{align}
a_r(q;p,e)
&=\frac{p}{w}-1
-q\left(\frac{2p}{3w}+1\right)
+\mathcal O(p^{-1},qp^{-1},q^2),
\label{eq:ar-fixed-frequency-p-e}
\\
x_t(q;p,e)
&=x-\frac{6xw}{p}
+q\,w\left[-\frac{2}{3}+\frac{31-28x}{6p}\right]
\nonumber\\
&\qquad
+\mathcal O(p^{-2},qp^{-2},q^2),
\label{eq:et-fixed-frequency-p-e}
\end{align}
extending Eqs.~\eqref{eq:ar-p-map}--\eqref{eq:et-e-map} to linear order in $q$.

At fixed $(p,e)$, we expand $\mathcal B$ in powers of $q$ as
\begin{equation}
\mathcal B(q;p,e)
=\mathcal B_0(p,e)+q\,\mathcal B_1(p,e)+\mathcal O(q^2).
\end{equation}
Substituting Eqs.~\eqref{eq:ar-fixed-frequency-p-e} and
\eqref{eq:et-fixed-frequency-p-e} yields
\begin{align}
\mathcal B_0={}& \frac{w^{13/2}}{p^{13/2}}F_0 + \frac{w^{15/2}}{p^{15/2}} \left(\frac{13}{2}F_0-6xF_0'-F_1\right)
\nonumber\\
&+\mathcal O(p^{-17/2}),
\label{eq:B0-p-e}
\\
\mathcal B_1={}&
\frac{w^{13/2}}{p^{13/2}}
\left(\frac{13}{3}F_0-\frac{2}{3}wF_0'\right)
\nonumber\\
&+\frac{w^{15/2}}{p^{15/2}}
\biggl[
39F_0+\frac{5-158x}{6}F_0'+4xwF_0''
\nonumber\\
&\quad \quad
+\frac{2}{3}wF_1'-5F_1-G_1
\biggr]
+\mathcal O(p^{-17/2}),
\label{eq:B1-p-e}
\end{align}
where $F_0$, $F_1$, and $G_1$, and their derivatives, are evaluated at $x=e^2$.

Similar to before, the prefactor in Eq.~\eqref{eq:Rtt-B} satisfies $(1+q)\nu^2 = q^2-3q^3+\mathcal O(q^4)$.  
The required expression at $\mathcal{O}(q^3)$ therefore becomes
\begin{equation}
\frac{\mathcal R_{\rm tt}}{m_2}
=\pi\left[
q^2\mathcal B_0
+q^3(\mathcal B_1-3\mathcal B_0)
\right]
+\mathcal O(q^4).
\label{eq:Rtt-q-expansion}
\end{equation}

Differentiating Eq.~\eqref{eq:Rtt-q-expansion} at fixed $(p,e)$, and matching to the expansion in Eq.~\eqref{eq:delta-z-tt-expansion}, gives
\begin{align}
\left.\frac{\partial}{\partial q}
\left(\frac{\mathcal R_{\rm tt}}{m_2}\right)
\right|_{p,e}
&=q\,\Delta z_{\rm tt}^{\rm 1SF}
+q^2\,\Delta z_{\rm tt}^{\rm 2SF}
+\mathcal O(q^3).
\label{eq:delta-z-tt-definition}
\end{align}
Matching powers of $q$ gives us the $1$SF and $2$SF contributions
\begin{align}
\Delta z_{\rm tt}^{\rm 1SF}
&=2\pi\mathcal B_0,
\\
\Delta z_{\rm tt}^{\rm 2SF}
&=3\pi(\mathcal B_1-3\mathcal B_0),
\label{eq:2SF-master}
\end{align}
where the $1$SF relation agrees with the result obtained in Eq.~\eqref{eq:z1SF-from-H}.

The contribution to the 2SF redshift can now be written in terms of the $5.5$PN and $6.5$PN contributions, 
\begin{align}
\Delta z_{\rm tt}^{\rm 2SF}
=\pi\left[
\frac{\mathcal Z_{11/2}^{(2)}(e)}{p^{13/2}}
+\frac{\mathcal Z_{13/2}^{(2)}(e)}{p^{15/2}}
\right]
+\mathcal O(p^{-17/2},e^{12}),
\label{eq:z2SF-tail-result}
\end{align}
where the coefficients are given by
\begin{align}
\mathcal Z_{11/2}^{(2)}
&=w^{13/2}\left(4F_0-2wF_0'\right),
\label{eq:Z11half-2SF}
\\
\mathcal Z_{13/2}^{(2)}
&=w^{15/2}\Bigl[
\frac{117}{2}F_0
+\frac{5}{2}(1-10x)F_0'
+12xwF_0''
\nonumber\\
&\qquad\quad
+2wF_1'-6F_1-3G_1
\Bigr].
\label{eq:Z13half-2SF}
\end{align}
Because the $2$SF expressions involve derivatives of the eccentricity, knowledge of $F_i$ through $\mathcal O(e^{12})$ is only sufficient to determine the $2$SF result through $\mathcal O(e^{10})$.

Finally, we can calculate the $2$SF inverse redshift invariant. 
We let $\Delta z^{\rm 1SF} = \Delta z_0^{\rm 1SF} +\Delta z_{\rm tt}^{\rm 1SF}$, where $\Delta z_0^{\rm 1SF}$ denotes the tail-free contribution.  
We can then linearise the cross-term appearing in Eq.~\eqref{eq:U-z-2SF} in the tail-of-tail interaction, resulting in
\begin{align}
\Delta U_{\rm tt}^{\rm 2SF} =-\frac{\Delta z_{\rm tt}^{\rm 2SF}}{z_{\rm geo}^2} +\frac{2\,\Delta z_0^{\rm 1SF}\, \Delta z_{\rm tt}^{\rm 1SF}}{z_{\rm geo}^3}.
\label{eq:U2SF-linear-tail}
\end{align}
Since $\Delta z_0^{\rm 1SF} = \mathcal O(p^{-1})$ and $\Delta z_{\rm tt}^{\rm 1SF} = \mathcal O(p^{-13/2})$, the cross term first contributes at $\mathcal O(p^{-15/2})$, i.e., at relative $1$PN order.
As such, we only need the following contributions from the tail-free terms~\cite{Cutler:1994pb,Akcay:2015pza,
Hopper:2015icj,Munna:2022gio}
\begin{align}
z_{\rm geo}
&=1-\frac{3}{2}\frac{w}{p}+\mathcal O(p^{-2}),
\\
\Delta z_0^{\rm 1SF}
&=\frac{w}{p}+\mathcal O(p^{-2}).
\end{align}
Substituting these into Eq.~\eqref{eq:U2SF-linear-tail} gives
\begin{equation}
\Delta U_{\rm tt}^{\rm 2SF}
=\pi\left[
\frac{\mathcal U_{11/2}^{(2)}(e)}{p^{13/2}}
+\frac{\mathcal U_{13/2}^{(2)}(e)}{p^{15/2}}
\right]
+\mathcal O(p^{-17/2},e^{12}),
\label{eq:U2SF-tail-result}
\end{equation}
Eccentricity expanding these coefficients leads to
\begin{align}
\mathcal U_{11/2}^{(2)}(e)={}&
\frac{417514}{1575}
+\frac{42586}{175}e^2
-\frac{35640523}{50400}e^4
\nonumber\\
&+\frac{804961}{32400}e^6
+\frac{10044408967}{96768000}e^8
\nonumber\\
&+\frac{216499199}{7680000}e^{10}
+\mathcal O(e^{12}),
\label{eq:U11half-2SF-result}
\\[1ex]
\mathcal U_{13/2}^{(2)}(e)={}&
-\frac{9851257}{7350}
-\frac{171446197}{39200}e^2
\nonumber\\
&+\frac{7769751313}{1411200}e^4
+\frac{117922725349}{40642560}e^6
\nonumber\\
&-\frac{4876131494921}{2709504000}e^8
\nonumber\\
&-\frac{171112938659393}{433520640000}e^{10}
+\mathcal O(e^{12}).
\label{eq:U13half-2SF-result}
\end{align}
Equations~\eqref{eq:U11half-2SF-result} and
\eqref{eq:U13half-2SF-result} give the conservative tail-of-tail contribution
to the 2SF inverse redshift through relative 1PN order and
$\mathcal O(e^{10})$, with terms quadratic in the tail-of-tail interaction entering beyond the PN accuracy considered here.  
The linearised result therefore gives the complete corresponding $2$SF inverse-redshift contributions at $5.5$PN and $6.5$PN.

We note that the circular-orbit limit of the $2$SF results derived above is not guaranteed to coincide with the corresponding quantity computed directly for circular orbits~\cite{Barack:2011ed,Akcay:2016dku}.
In particular, they are defined by different comparisons.  
The standard circular-orbit self-force calculation compares circular configurations in the background and perturbed spacetimes at the same azimuthal frequency $\Omega_\phi$.
In contrast, the eccentric-orbit invariants derived above compare configurations with the same fundamental frequencies $(\widehat\Omega_r,\widehat\Omega_\phi)$~\cite{Akcay:2016dku}.
Using fixed frequencies, the perturbed orbit is generically non-circular, even when the reference geodesic is circular.
As such, the two limits are expected to differ by terms proportional to the $\mathcal{O}(q)$ periastron advance~\cite{Barack:2011ed,Akcay:2015pza,Akcay:2016dku}.
We also note that the $1$SF result does agree with the circular-orbit calculation, as the offset between the two comparisons is proportional to the radial action of the reference geodesic, which vanishes for circular orbits thanks to the first law~\cite{LeTiec:2011ab}.

\subsection{Cross-validation with EOB}
In this section, we reproduce the redshift coefficients directly from the linearised EOB Hamiltonian, providing a consistency check of both the fixed-frequency expansion and the matching to the tail-of-tail EOB potentials~\cite{Damour:2009sm}.  
As discussed above, the fixed-frequency expansion lowers the available eccentricity order by two, so an EOB Hamiltonian through $\mathcal O(e_t^4)$ determines the redshift through $\mathcal O(e^2)$.
Here, we work to $\mathcal{O}(e^4_t)$ for computational convenience, though this comparison can be trivially extended to higher orders.

Our starting point is to linearise the effective Hamiltonian in the tail-of-tail potentials~\cite{Bini:2025zvk}, giving
\begin{align}
\frac{H^{\rm tt}_{\rm EOB}}{M} =\frac{\nu}{2\hat H_{\rm eff}\hat H}
\Bigl[&\delta A\left(1+p_\phi^2u^2\right)
\nonumber\\
&+p_r^2\left(2A\bar D\,\delta A+A^2\,\delta\bar D\right)
+A\,\delta Q\Bigr],
\label{eq:Htt-EOB-linear}
\end{align}
where $u=GM/r_{\rm EOB}$.  Through $\mathcal O(e_t^4)$, only $\delta A$, $\delta\bar D$, and the $p_r^4$ sector of $\delta Q$ contribute.
The required EOB coefficients are therefore $\{a_{6.5},a_{7.5},\bar d_{5.5},\bar d_{6.5},q_{4,4.5},q_{4,5.5}\}$.
Evaluating Eq.~\eqref{eq:Htt-EOB-linear} along the 1PN QK orbit in EOB coordinates, and averaging over one radial period, yields
\begin{align}
\left\langle H^{\rm tt}_{\rm EOB}\right\rangle_{\eta^0}
&=\frac{\pi\nu^2}{a_r^{{\rm e}\,13/2}}
\left[
\frac{6848}{525}
+\frac{49969}{315}e_t^{{\rm e}2}
+\frac{919237}{1260}e_t^{{\rm e}4}
\right],
\label{eq:HEOB-LO}
\\
\left\langle H^{\rm tt}_{\rm EOB}\right\rangle_{\eta^2}
&=-\frac{\pi\nu^2}{a_r^{{\rm e}\,15/2}}
\Biggl[
\frac{368693}{7350}+\frac{4982}{315}\nu
\nonumber\\
&\qquad
-\left(\frac{3982}{45}-\frac{6263599}{22050}\nu\right)e_t^{{\rm e}2}
\nonumber\\
&\qquad
-\left(\frac{355973869}{94080}
-\frac{208757957}{117600}\nu\right)e_t^{{\rm e}4}
\Biggr],
\label{eq:HEOB-NLO}
\end{align}
where $a_r^{\rm e}=a_r+\eta^2$ and
$e_t^{\rm e}=e_t(1-\eta^2\nu/a_r)$.

The EOB fixed-frequency map can be written in terms of $(x,p,w)$ as 
\begin{align}
a_r^{\rm e}(q;p,e) &=\frac{p}{w} -q\left(\frac{2p}{3w}+1\right) +\mathcal O(p^{-1},qp^{-1},q^2),
\label{eq:arE-fixed-frequency}
\\
x_t^{\rm e}(q;p,e) &=x-\frac{6xw}{p} +q\,w\left[-\frac{2}{3}+\frac{31-40x}{6p}\right]
\nonumber\\
&\qquad
+\mathcal O(p^{-2},qp^{-2},q^2).
\label{eq:xtE-fixed-frequency}
\end{align}
We now iterate the same procedure as above to determine the redshift invariant and its inverse. 
Following this procedure gives
\begin{align}
\Delta z^{\rm 1SF}_{\rm tt}
={}&\pi\left(
\frac{13696}{525}+\frac{232618}{1575}e^2
\right)\frac{1}{p^{13/2}}
\nonumber\\
&-\pi\left(
\frac{368693}{3675}+\frac{4296083}{4410}e^2
\right)\frac{1}{p^{15/2}},
\label{eq:z1SF-e2}
\\
\Delta z^{\rm 2SF}_{\rm tt}
={}&-\pi\left(
\frac{417514}{1575}+\frac{42586}{175}e^2
\right)\frac{1}{p^{13/2}}
\nonumber\\
&+\pi\left(
\frac{16079941}{7350}+\frac{1605811933}{352800}e^2
\right)\frac{1}{p^{15/2}},
\label{eq:z2SF-e2}
\end{align}
and the corresponding inverse-redshift contributions by
\begin{align}
\Delta U^{\rm 1SF}_{\rm tt}
={}&-\pi\left(
\frac{13696}{525}+\frac{232618}{1575}e^2
\right)\frac{1}{p^{13/2}}
\nonumber\\
&+\pi\left(
\frac{81077}{3675}+\frac{2687231}{4410}e^2
\right)\frac{1}{p^{15/2}},
\label{eq:U1SF-e2}
\\
\Delta U^{\rm 2SF}_{\rm tt}
={}&\pi\left(
\frac{417514}{1575}+\frac{42586}{175}e^2
\right)\frac{1}{p^{13/2}}
\nonumber\\
&-\pi\left(
\frac{9851257}{7350}+\frac{171446197}{39200}e^2
\right)\frac{1}{p^{15/2}}.
\label{eq:U2SF-e2}
\end{align}
These expressions agree with Eqs.~\eqref{eq:U11half-result},
\eqref{eq:U13half-result}, \eqref{eq:U11half-2SF-result}, and
\eqref{eq:U13half-2SF-result} through $\mathcal O(e^2)$.  
This provides a useful consistency check on the fixed-frequency expansion and matching, including the $\mathcal O(\nu^2)$ contributions to $a_{7.5}$, $\bar d_{6.5}$, and $q_{4,5.5}$.

\section{Concluding Remarks}
We have extended the conservative tail-of-tail dynamics of eccentric, non-spinning binaries through relative $1$PN order and $\mathcal O(e_t^{12})$.
Matching the Delaunay-averaged Hamiltonian to the EOB dynamics determines the $5.5$PN and $6.5$PN contributions to the non-geodesic $Q$ potential through the $p_r^{12}$ sector, including their complete dependence on the symmetric mass ratio. 
The terms linear in $\nu$ reproduce the available first-order self-force information, while the quadratic terms constitute qualitatively new eccentric 2SF predictions.

These results are supported by several complementary calculations. 
The time-split derivation of the averaged Hamiltonian agrees with its spectral Fourier--Bessel representation, and the first-law calculation reproduces the known 1SF redshift through $\mathcal O(e^{12})$~\cite{Munna:2022gio}. 
At 2SF order, the fixed-frequency expansion yields the tail-of-tail contribution to the inverse redshift at $5.5$PN and $6.5$PN through $\mathcal O(e^{10})$, with a direct EOB calculation providing an internal consistency check through $\mathcal O(e^{2})$.

Three extensions of this work are of particular interest. 
First, direct eccentric 2SF calculations would provide an independent test of the new weak-field predictions and connect them to the strong-field regime. 
Second, the higher-$p_r$ sectors of the $Q$ potential obtained here could be analytically continued to unbound orbits. 
Reconstructing the corresponding tail-of-tail scattering angle would allow a direct bound--unbound comparison with the scattering calculation of Ref.~\cite{Bini:2025zvk}, and would help clarify the role of the non-geodesic terms in high-PM scattering dynamics. 
Such predictions could also be incorporated into resummed EOB models of the scattering dynamics~\cite{Buonanno:2024vkx,Damour:2025uka}, whose gauge flexibility was recently explored in Ref.~\cite{Clark:2025kvu}, and confronted directly against numerical-relativity scattering simulations~\cite{Damour:2014afa,Rettegno:2023ghr,
Swain:2024ngs,Albanesi:2024xus,Long:2025nmj,Clark:2026bgg}.
Finally, extending the analysis beyond fractional $1$PN accuracy would determine the half-integer contributions beyond $6.5$PN, at the cost of higher-PN source multipoles and orbital dynamics, and help to further constrain the eccentric EOB potentials.

\begin{acknowledgments}
G.P. thanks Patricia Schmidt for useful discussions throughout this project.
G.P. is very grateful for support from a Royal Society University Research Fellowship URF{\textbackslash}R1{\textbackslash}221500 and RF{\textbackslash}ERE{\textbackslash}221015. 
G.P. acknowledges support from STFC grants ST/V005677/1 and UKRI2493, and the UKSA grant UKRI/ST/B000971/1. 

\end{acknowledgments}

\appendix
\section{Multipole Moment Bilinears}
\label{app:bilinears}
\begin{table*}  
\caption{\label{tab:1} Multipole-moment bilinears for $I^{(3,4)}_2 (u, u')$ up to fractional $1$PN order and $\mathcal{O}(e^4_t)$. The complete expressions up to $\mathcal{O}(e^{12}_t)$ can be found in the ancillary material.  }
\renewcommand{\arraystretch}{1.12}
\begin{ruledtabular}
\begin{tabular}{l|l}
$I_{2(\eta^0,e_t^0)}^{(3,4)}(u,u')$ & $\frac{64\nu^2}{a_r^{13/2}}\left[\sin(2(u-u'))\right]$\\
$I_{2(\eta^0,e_t^1)}^{(3,4)}(u,u')$ & $\frac{4\nu^2}{a_r^{13/2}}\left[38\sin(3u-2u')+13\sin(2u-u')+65\sin(2u-3u')+10\sin(u-2u')\right]$\\
$I_{2(\eta^0,e_t^2)}^{(3,4)}(u,u')$ & $\frac{\nu^2}{6a_r^{13/2}}\left[-120\sin(2u')+1320\sin(4u-2u')+741\sin(3u-u')+3705\sin(3(u-u'))\right.$\\
& $\left.+2280\sin(2(u-u'))+3420\sin(2u-4u')+197\sin(u-u')+975\sin(u-3u')+156\sin(2u) + 2\sin(u+u')\right]$\\
$I_{2(\eta^0,e_t^3)}^{(3,4)}(u,u')$ & $\frac{\nu^2}{12a_r^{13/2}}\left[-195\sin(u')-975\sin(3u')+3000\sin(5u-2u')+2145\sin(4u-u')+10725\sin(4u-3u')\right.$\\
& $\left.+10206\sin(3u-2u')+16245\sin(3u-4u')+3282\sin(2u-u')+14340\sin(2u-3u')+10920\sin(2u-5u')\right.$\\
& $\left.+2864\sin(u-2u')+4275\sin(u-4u')+199\sin(u)+741\sin(3u)-106\sin(u+2u')+102\sin(2u+u')\right]$\\
$I_{2(\eta^0,e_t^4)}^{(3,4)}(u,u')$ & $\frac{\nu^2}{24a_r^{13/2}}\left[-2970\sin(2u')-4275\sin(4u')+5880\sin(6u-2u')+4875\sin(5u-u')+24375\sin(5u-3u')\right.$\\
& $\left.+29790\sin(4u-2u')+47025\sin(4(u-u'))+14559\sin(3u-u')+63045\sin(3(u-u'))\right.$\\
& $\left.+51870\sin(3u-5u')+29562\sin(2(u-u'))+59100\sin(2u-4u')+28560\sin(2u-6u')+4137\sin(u-u')\right.$\\
& $\left.+17975\sin(u-3u')+13650\sin(u-5u')+3384\sin(2u)+2145\sin(4u)-33\sin(u+u')-925\sin(u+3u')\right.$\\
& $\left.-102\sin(2u+2u')+468\sin(3u+u')\right]$\\
\hline
$I_{2(\eta^2,e_t^0)}^{(3,4)}(u,u')$ & $\frac{32\nu^2}{21a_r^{15/2}}\left[252(u-u')\cos(2(u-u'))-\left(487-201\nu\right)\sin(2(u-u'))\right]$\\
$I_{2(\eta^2,e_t^1)}^{(3,4)}(u,u')$ & $\frac{2\nu^2}{21a_r^{15/2}}\left[9576(u-u')\cos(3u-2u')+3276(u-u')\cos(2u-u')+16380(u-u')\cos(2u-3u')\right.$\\
& $\left.+2520(u-u')\cos(u-2u')+2(3657\nu-6805)\sin(3u-2u')+7(393\nu-919)\sin(2u-u')\right.$\\
& $\left.+7(1797\nu-3815)\sin(2u-3u')+2(1143\nu-3167)\sin(u-2u')\right]$\\
$I_{2(\eta^2,e_t^2)}^{(3,4)}(u,u')$ & $\frac{\nu^2}{84a_r^{15/2}}\left[10080(u-u')\cos(2u')+110880(u-u')\cos(4u-2u')+62244(u-u')\cos(3u-u')\right.$\\
& $\left.+311220(u-u')\cos(3(u-u'))+223776(u-u')\cos(2(u-u'))+287280(u-u')\cos(2u-4u')\right.$\\
& $\left.+16380(u-u')\cos(u-u')+81900(u-u')\cos(u-3u')+13104(u-u')\cos(2u)-8(1143\nu-2915)\sin(2u')\right.$\\
& $\left.+40(2001\nu-2641)\sin(4u-2u')-\left(90403-50163\nu\right)\sin(3u-u')\right.$\\
& $\left.-\left(348275-228471\nu\right)\sin(3(u-u'))+24(7235\nu-13971)\sin(2(u-u'))\right.$\\
& $\left.+4(52599\nu-93385)\sin(2u-4u')-\left(42359-15595\nu\right)\sin(u-u')\right.$\\
& $\left.+5(14373\nu-36221)\sin(u-3u')+28(393\nu-739)\sin(2u)+2(23\nu-339)\sin(u+u')\right]$\\
$I_{2(\eta^2,e_t^3)}^{(3,4)}(u,u')$ & $\frac{\nu^2}{168a_r^{15/2}}\left[16380(u-u')\cos(u')+81900(u-u')\cos(3u')+252000(u-u')\cos(5u-2u')+180180(u-u')\cos(4u-u')\right.$\\
& $\left.+900900(u-u')\cos(4u-3u')+1010520(u-u')\cos(3u-2u')+1364580(u-u')\cos(3u-4u')\right.$\\
& $\left.+327600(u-u')\cos(2u-u')+1466640(u-u')\cos(2u-3u')+917280(u-u')\cos(2u-5u')\right.$\\
& $\left.+279720(u-u')\cos(u-2u')+359100(u-u')\cos(u-4u')+16380(u-u')\cos(u)+62244(u-u')\cos(3u)\right.$\\
& $\left.+10080(u-u')\cos(u+2u')+8064(u-u')\cos(2u+u')+\left(38405-15549\nu\right)\sin(u')\right.$\\
& $\left.-5(14373\nu-32945)\sin(3u')+8(21165\nu-16127)\sin(5u-2u')+35(3933\nu-5065)\sin(4u-u')\right.$\\
& $\left.+35(17817\nu-16745)\sin(4u-3u')+2(373053\nu-498431)\sin(3u-2u')\right.$\\
& $\left.+133(7167\nu-8095)\sin(3u-4u')+14(18713\nu-34367)\sin(2u-u')\right.$\\
& $\left.+4(271311\nu-426346)\sin(2u-3u')+56(11316\nu-15805)\sin(2u-5u')\right.$\\
& $\left.+12(20178\nu-47039)\sin(u-2u')+5(60465\nu-135109)\sin(u-4u')\right.$\\
& $\left.+\left(15641\nu-36737\right)\sin(u)+\left(50163\nu-66463\right)\sin(3u)\right.$\\
& $\left.-2(3219\nu-2599)\sin(u+2u')+2(3881\nu-5097)\sin(2u+u')\right]$\\
$I_{2(\eta^2,e_t^4)}^{(3,4)}(u,u')$ & $\frac{\nu^2}{336a_r^{15/2}}\left[16380(u-u')+289800(u-u')\cos(2u')+359100(u-u')\cos(4u')+493920(u-u')\cos(6u-2u')\right.$\\
& $\left.+409500(u-u')\cos(5u-u')+2047500(u-u')\cos(5u-3u')+2945880(u-u')\cos(4u-2u')\right.$\\
& $\left.+3950100(u-u')\cos(4(u-u'))+1470924(u-u')\cos(3u-u')+6540660(u-u')\cos(3(u-u'))\right.$\\
& $\left.+4357080(u-u')\cos(3u-5u')+3374784(u-u')\cos(2(u-u'))+6113520(u-u')\cos(2u-4u')\right.$\\
& $\left.+2399040(u-u')\cos(2u-6u')+409500(u-u')\cos(u-u')+1833300(u-u')\cos(u-3u')\right.$\\
& $\left.+1146600(u-u')\cos(u-5u')+335664(u-u')\cos(2u)+180180(u-u')\cos(4u)+26460(u-u')\cos(u+u')\right.$\\
& $\left.+81900(u-u')\cos(u+3u')+8064(u-u')\cos(2u+2u')+38304(u-u')\cos(3u+u')\right.$\\
& $\left.-2(124287\nu-261901)\sin(2u')-25(12093\nu-24149)\sin(4u')+56(5415\nu-797)\sin(6u-2u')\right.$\\
& $\left.-\left(222401-292395\nu\right)\sin(5u-u')+5(262995\nu-85901)\sin(5u-3u')\right.$\\
& $\left.+2(1030845\nu-799819)\sin(4u-2u')+65(39903\nu-19475)\sin(4(u-u'))\right.$\\
& $\left.-\left(1396449-1118377\nu\right)\sin(3u-u')-\left(4241879-4584735\nu\right)\sin(3(u-u'))\right.$\\
& $\left.+14(204474\nu-141175)\sin(3u-5u')+154(16471\nu-25199)\sin(2(u-u'))\right.$\\
& $\left.+20(215991\nu-259772)\sin(2u-4u')+56(27561\nu-27016)\sin(2u-6u')\right.$\\
& $\left.-\left(815827-363595\nu\right)\sin(u-u')+5(302157\nu-607141)\sin(u-3u')\right.$\\
& $\left.+70(13110\nu-25321)\sin(u-5u')+4(67436\nu-98893)\sin(2u)+35(3933\nu-3085)\sin(4u)\right.$\\
& $\left.+\left(-59\nu-14041\right)\sin(u+u')-35(1467\nu-1117)\sin(u+3u')\right.$\\
& $\left.+14(329\nu-2801)\sin(2u+2u')+2(17597\nu-11622)\sin(3u+u')\right]$\\
\end{tabular}
\end{ruledtabular}
\end{table*}

\begin{table*}  
\caption{\label{tab:2} Multipole-moment bilinears for $J^{(3,4)}_2 (u, u')$ up to fractional $1$PN order and $\mathcal{O}(e^4_t)$. The complete expressions up to $\mathcal{O}(e^{12}_t)$ can be found in the ancillary material.}
\renewcommand{\arraystretch}{1.175}
\begin{ruledtabular}
\begin{tabular}{l|l}
$J_{2(\eta^0,e_t^0)}^{(3,4)}(u,u')$ & $\frac{\nu^2}{2a_r^{15/2}}\left(1-4\nu\right)\left[\sin(u-u')\right]$\\
$J_{2(\eta^0,e_t^1)}^{(3,4)}(u,u')$ & $\frac{\nu^2}{4a_r^{15/2}}\left(1-4\nu\right)\left[-\sin(u')+7\sin(2u-u')+15\sin(u-2u')+\sin(u)\right]$\\
$J_{2(\eta^0,e_t^2)}^{(3,4)}(u,u')$ & $\frac{\nu^2}{8a_r^{15/2}}\left(1-4\nu\right)\left[-15\sin(2u')+25\sin(3u-u')+105\sin(2(u-u'))+38\sin(u-u')+90\sin(u-3u')+7\sin(2u)\right.$\\
& $\left.+\sin(u'+u)\right]$\\
$J_{2(\eta^0,e_t^3)}^{(3,4)}(u,u')$ & $\frac{\nu^2}{16a_r^{15/2}}\left(1-4\nu\right)\left[-37\sin(u')-90\sin(3u')+65\sin(4u-u')+375\sin(3u-2u')+243\sin(2u-u')\right.$\\
& $\left.+630\sin(2u-3u')+390\sin(u-2u')+350\sin(u-4u')+39\sin(u)+25\sin(3u)+5\sin(u+2u')\right.$\\
& $\left.+9\sin(2u+u')\right]$\\
$J_{2(\eta^0,e_t^4)}^{(3,4)}(u,u')$ & $\frac{\nu^2}{32a_r^{15/2}}\left(1-4\nu\right)\left[-385\sin(2u')-350\sin(4u')+140\sin(5u-u')+975\sin(4u-2u')+860\sin(3u-u')\right.$\\
& $\left.+2250\sin(3(u-u'))+2385\sin(2(u-u'))+2450\sin(2u-4u')+702\sin(u-u')+2120\sin(u-3u')\right.$\\
& $\left.+1050\sin(u-5u')+252\sin(2u)+65\sin(4u)+36\sin(u+u')+20\sin(u+3u')+65\sin(2u+2u')\right.$\\
& $\left.+30\sin(3u+u')\right]$\\
\end{tabular}
\end{ruledtabular}
\end{table*}

\begin{table*}  
\caption{\label{tab:3} Multipole-moment bilinears for $I^{(4,5)}_3 (u, u')$ up to fractional $1$PN order and $\mathcal{O}(e^4_t)$. The complete expressions up to $\mathcal{O}(e^{12}_t)$ can be found in the ancillary material.}
\renewcommand{\arraystretch}{1.175}
\begin{ruledtabular}
\begin{tabular}{l|l}
$I_{3(\eta^0,e_t^0)}^{(4,5)}(u,u')$ & $\frac{3\nu^2}{20a_r^{15/2}}\left(1-4\nu\right)\left[32805\sin(3(u-u'))+\sin(u-u')\right]$\\
$I_{3(\eta^0,e_t^1)}^{(4,5)}(u,u')$ & $\frac{3\nu^2}{40a_r^{15/2}}\left(1-4\nu\right)\left[-\sin(u')+212625\sin(4u-3u')+59535\sin(3u-2u')+316305\sin(3u-4u')-17\sin(2u-u')\right.$\\
& $\left.+40095\sin(2u-3u')-33\sin(u-2u')+\sin(u)\right]$\\
$I_{3(\eta^0,e_t^2)}^{(4,5)}(u,u')$ & $\frac{3\nu^2}{80a_r^{15/2}}\left(1-4\nu\right)\left[33\sin(2u')+801900\sin(5u-3u')+385875\sin(4u-2u')+2050125\sin(4(u-u'))\right.$\\
& $\left.+73621\sin(3u-u')+932310\sin(3(u-u'))+1652805\sin(3u-5u')+73326\sin(2(u-u'))\right.$\\
& $\left.+386595\sin(2u-4u')-82\sin(u-u')+42225\sin(u-3u')-17\sin(2u)-11\sin(u+u')\right]$\\
$I_{3(\eta^0,e_t^3)}^{(4,5)}(u,u')$ & $\frac{3\nu^2}{160a_r^{15/2}}\left(1-4\nu\right)\left[71\sin(u')-42225\sin(3u')+2296350\sin(6u-3u')+1455300\sin(5u-2u')\right.$\\
& $\left.+7731900\sin(5u-4u')+477485\sin(4u-u')+5533350\sin(4u-3u')+10712625\sin(4u-5u')\right.$\\
& $\left.+1457607\sin(3u-2u')+6714915\sin(3u-4u')+6245505\sin(3u-6u')+90879\sin(2u-u')\right.$\\
& $\left.+1137435\sin(2u-3u')+2020095\sin(2u-5u')+77283\sin(u-2u')+408665\sin(u-4u')\right.$\\
& $\left.-93\sin(u)+73621\sin(3u)+161\sin(u+2u')+69\sin(2u+u')\right]$\\
$I_{3(\eta^0,e_t^4)}^{(4,5)}(u,u')$ & $\frac{3\nu^2}{320a_r^{15/2}}\left(1-4\nu\right)\left[-77122\sin(2u')-408665\sin(4u')+5528250\sin(7u-3u')+4167450\sin(6u-2u')\right.$\\
& $\left.+22141350\sin(6u-4u')+1801190\sin(5u-u')+20727000\sin(5u-3u')\right.$\\
& $\left.+40401900\sin(5(u-u'))+8512695\sin(4u-2u')+38610975\sin(4(u-u'))\right.$\\
& $\left.+40480125\sin(4u-6u')+1780532\sin(3u-u')+13860930\sin(3(u-u'))\right.$\\
& $\left.+32193420\sin(3u-5u')+19144755\sin(3u-7u')+1783074\sin(2(u-u'))\right.$\\
& $\left.+8161250\sin(2u-4u')+7633395\sin(2u-6u')+95742\sin(u-u')+1233860\sin(u-3u')\right.$\\
& $\left.+2138115\sin(u-5u')+90948\sin(2u)+477485\sin(4u)+324\sin(u+u')-40855\sin(u+3u')\right.$\\
& $\left.+1157\sin(2u+2u')+59949\sin(3u+u')\right]$\\
\end{tabular}
\end{ruledtabular}
\end{table*}

\clearpage
\section{Fourier-Bessel Amplitudes: A Worked Example}
\label{app:NewtonianMassQuadrupole}
Here we schematically outline the derivation of $D_p = \mathcal{A}_p [\rho^2]$, where $\rho = 1 - e_t \cos u$.
As was detailed in Sec.~\ref{sec:spectral}, the Fourier-Bessel amplitude can be written as 
\begin{align}
    D_p &= \mathcal{A}_p \left[ \rho^2 \right] = \frac{1}{2 \pi} \int^{2 \pi}_0 du (1 - e_t \cos u) \rho^2 e^{-i p (u - e_t \sin u)},
\end{align}
which can also be expressed in terms of a Laurent expansion $\rho(u) F(u) = \sum_j c_j z^j$, with $z = e^{iu}$. 
Here, the weight function $\rho(u) F(u)$ is trivially given by $\rho^3 = (1 - e_t \cos u)^3$. 
Writing $\cos u = (z + z^{-1})/2$, the weight function can be expanded as
\begin{align}
    \rho^3 &= \left[1 - \frac{e_t}{2} (z + z^{-1}) \right]^3, \\
    &= c_0 + \displaystyle\sum_{j=1}^{3} c_j (z^j + z^{-j}),
\end{align}
where the expansion terminates at $|j| = 3$ due to the trigonometric polynomial nature of the expressions. 
The resulting coefficients can be matched, leading to
\begin{align*}
c_0 &= 1 + \frac{3}{2} e^2_t,\\
c_1 &= -\frac{3}{2} e_t \left(1 + \frac{e^2_t}{4} \right), \\
c_2 &= \frac{3}{4} e^2_t, \\
c_3 &= -\frac{e^3_t}{8}.
\end{align*}
This allows us to express $D_p$ as
\begin{align}
    D_p &= c_0 J_p + c_1 (J_{p-1} + J_{p+1}) + c_2 (J_{p-2} + J_{p+2}) \\ \nonumber & \qquad \qquad \qquad \qquad + c_3 (J_{p-3} + J_{p+3}),
\end{align}
where $J_m = J_m(p e_t)$. 
Using the standard Bessel recurrence relations,
\begin{align}
    J_{n-1}(p e_t) + J_{n+1}(p e_t) &= \frac{2n}{p e_t} J_n(p e_t), \\
    J_{n-1}(p e_t) - J_{n+1}(p e_t) &= 2 J_n'(p e_t),
\end{align}
where prime denotes differentiation with respect to the argument, we find that, for $p \neq 0$, $D_p$ reduces to
\begin{align}
    D_p = -\frac{2}{p^2} J_p(p e_t),
\end{align}
in agreement with Eq.~\eqref{eq:Dp}. 
The derivation of the other coefficients follows a similar pattern.

\section{Fourier-Bessel Amplitudes: $U_p^{\pm}$ and $V_p^{\pm}$}
\label{app:bessel-amplitudes}
The Fourier amplitudes $U_p^{\pm}$ and $V_p^{\pm}$ introduced in Eq.~\eqref{eq:UpVp} are given by
\begin{align}
U_p^\pm
&=\frac{3}{e_t^3p^3}
\left[
\pm s A_p^\pm J_p(pe_t)
+e_t B_p^\pm J_p'(pe_t)
\right],
\\
V_p^\pm
&=\frac{6\mp5ps}{p^3}
\left[
J_p'(pe_t)
\pm\frac{s}{e_t}J_p(pe_t)
\right],
\end{align}
where $A_p^{\pm}$ and $B_p^{\pm}$ are polynomials in $p$ and $s$, 
\begin{align}
A_p^\pm
&\equiv 4p^2s^4+2s^2+6
\mp ps(7+5s^2),
\\
B_p^\pm
&\equiv 4p^2s^4+6s^2+2
\mp ps(5+7s^2).
\end{align}

\clearpage 
\bibliography{references.bib}

\end{document}